\documentclass{ieeetj}
\usepackage{cite}
\usepackage{bm}
\usepackage{algorithmic}
\usepackage{graphicx,color}
\usepackage{xcolor}
\usepackage{textcomp}
 \usepackage{amsmath,amsfonts,amssymb,amsthm}
\usepackage{bm}
\usepackage{xcolor}
\usepackage{hyperref}
\hypersetup{hidelinks=true}
\usepackage{algorithm,algorithmic}
\usepackage{tikz}
\usepackage{placeins}
\usetikzlibrary{arrows.meta,positioning,shapes.geometric}
\usepackage[final]{changes}
\def\BibTeX{{\rm B\kern-.05em{\sc i\kern-.025em b}\kern-.08em
    T\kern-.1667em\lower.7ex\hbox{E}\kern-.125emX}}
\AtBeginDocument{\definecolor{tmlcncolor}{cmyk}{0.93,0.59,0.15,0.02}\definecolor{NavyBlue}{RGB}{0,86,125}}

\newtheorem{assumption}{Assumption}

\def\OJlogo{\vspace{-4pt}$<$Society logo(s) and publication title will appear here.$>$}
\def\seclogo{\vspace{10pt}$<$Society logo(s) and publication title will appear here.$>$}

\usetikzlibrary{patterns}
\tikzstyle{startstop} = [rectangle, rounded corners, 
minimum width=0.5cm, 
minimum height=0.5cm,
text centered, 
draw=black, 
fill=orange!30] 
\tikzstyle{io} = [trapezium, 
trapezium stretches=true, 
trapezium left angle=70, 
trapezium right angle=110, 
minimum width=0.5cm, 
minimum height=0.5cm, text centered, 
draw=black, fill=blue!30]
\tikzstyle{process} = [rectangle, 
minimum width=0.5cm, 
minimum height=0.5cm, 
text centered, 
text width=3.2cm, 
draw=black, 
fill=green!30] 
\tikzstyle{decision} = [diamond, 
minimum width=0.1cm, 
minimum height=0.1cm, 
text centered, 
draw=black, 
fill=green!30]
\tikzstyle{arrow} = [thick,->,>=stealth]

\title{Modeling, Optimization and Electromagnetic Validation of Stacked Intelligent Metasurfaces by Using a Multiport Network Model}
\author{
Giuseppe Pettanice\IEEEauthorrefmark{1},
Andrea Abrardo\IEEEauthorrefmark{2},
Alberto Toccafondi\IEEEauthorrefmark{2},
Marco Di Renzo\IEEEauthorrefmark{3}\\[6pt]
\IEEEauthorrefmark{1}Department of Information Engineering, Computer Science and Mathematics, Centre Ex-EMERGE, University of L'Aquila, L'Aquila, Italy (e-mail:giuseppe.pettanice@guest.univaq.it)\\
\IEEEauthorrefmark{2}Department of Information Engineering and Mathematics (DIISM), 
University of Siena, Italy and Consorzio Nazionale Interuniversitario Telecomunicazioni (CNIT),  
(e-mail: abrardo@unisi.it; alberto.toccafondi@unisi.it)\\
\IEEEauthorrefmark{3} CNRS, CentraleSup\'elec, Institute of Electronics and Digital Technologies, Avenue de la Boulaie, 35576 Cesson-S\'evign\'e, France,(marco.direnzo@centralesupelec.fr \\ King's College London, Department of Engineering - Centre for Telecommunications Research, WC2R 2LS London, United Kingdom. (marco.di\_renzo@kcl.ac.uk)
}
\date{}

\begin{document}
\receiveddate{XX Month, XXXX}
\reviseddate{XX Month, XXXX}
\accepteddate{XX Month, XXXX}
\publisheddate{XX Month, XXXX}
\currentdate{XX Month, XXXX}
\doiinfo{XXXX.2022.1234567}


\begin{abstract}
Stacked intelligent metasurfaces (SIMs) extend the concept of reconfigurable intelligent surfaces by cascading multiple programmable layers, enabling advanced electromagnetic wave transformations for communication and sensing applications. However, most existing optimization frameworks rely on simplified channel abstractions that may overlook key electromagnetic effects such as multiport coupling, circuit losses, and non-ideal hardware behavior. In this paper, we develop a modeling and optimization framework for SIMs based on a multiport network representation using scattering parameters. The proposed formulation captures realistic circuit characteristics and mutual interactions among SIM ports while remaining amenable to optimization. The resulting models are validated through electromagnetic simulations, enabling a systematic comparison between idealized and practical SIM configurations. Numerical results for communication and sensing scenarios confirm that the proposed framework provides accurate performance predictions and enables the effective design of SIM configurations under realistic electromagnetic conditions.
\end{abstract}

\maketitle
\section{Introduction}
\label{sec:intro}
Reconfigurable intelligent surfaces (RISs) and, more broadly, smart radio
environments have emerged as a promising paradigm to endow the wireless
propagation channel with software-defined and programmable features
\cite{DiRenzo:19b,DiRenzo2020_JSAC,RenzoDT22,WuZhaZheYouZha:20,DBLP:journals/vtm/BasarALWJYDS24}.
Within this vision, stacked intelligent metasurfaces (SIMs) extend the RIS
concept by cascading multiple reconfigurable layers, thus enabling richer wave
transformations and advanced functionalities in the electromagnetic domain, such
as holographic Multiple Input Multiple Output (MIMO), wave-domain multiuser beamforming, near-field beamfocusing,
and sensing functionalities
\cite{DBLP:journals/jsac/AnXNAHYH23,DBLP:journals/ojcs/HassanARDY24,DBLP:conf/icc/AnRDY23,DBLP:conf/vtc/JiaALGRDY24,DBLP:journals/wcl/NiuAPGCD24,An_2024_ref4}.

A key motivation behind SIMs is their potential to realize a form of
\emph{electromagnetic signal processing} directly in the wave domain. This
perspective resonates with diffractive and metasurface-based analog computing
architectures \cite{lin2018all,DiffractiveDNN}, and motivates using SIMs as
hardware platforms for wave-domain linear transformations.

However, exploiting SIMs for such advanced processing requires models that are
both (i) sufficiently accurate from an electromagnetic standpoint (e.g.,
including mutual coupling and multi-port interactions) and (ii) sufficiently
tractable to support systematic optimization.
Most of the literature on RIS/SIM optimization adopts simplified end-to-end
communication models, which may hide critical electromagnetic effects and can
lead to performance gaps when moving from theory to practical implementations.

Recent works have started addressing electromagnetic consistency by developing
mutual-coupling aware models and optimization methods for RISs
\cite{DR1,DR2,ABR_MUTUAL,RafiqueHZNMRDY23,AbeywickramaZWY20}, and by proposing
multiport network formulations based on scattering parameters
\cite{abrardo_E}.
For stacked architectures, a comprehensive multiport model and optimization
framework has been introduced in \cite{Abra_SIM1}, while physically consistent
models for SIMs implemented with beyond diagonal RIS architectures have been
recently discussed in \cite{Nerini_Clerckx_SIM,Li_2023_ref1,10155675}. Despite
this progress, there remains a gap between optimization-friendly abstractions
and the electromagnetic behavior observed in realistic implementations,
especially when tunable circuits, losses, and coupling
effects are explicitly accounted for, as predicted by full-wave
electromagnetic simulations.

The present paper contributes to this research direction by pushing forward
electromagnetically consistent modeling and optimization with a special focus on
\emph{model verification through electromagnetic simulations}.
Specifically, starting from a multiport network-theoretic formulation, we
derive an S-parameter representation tailored to SIM optimization that can
incorporate non-ideal circuits behavior, and validate the resulting model
and optimized configuration through electromagnetic simulations.
Ultimately, this work aims to provide a baseline framework that bridges
optimization-friendly abstractions and electromagnetically accurate models,
supporting the development of optimization methods for \emph{practical} SIM
implementations.
\section{Contributions}
\label{sec:contributions}
This work proposes multiport network-theoretic SIM models that preserve electromagnetic consistency while keeping the analytical complexity tractable for optimization in communication and sensing scenarios. Unlike many existing studies based on highly idealized abstractions, the proposed framework explicitly accounts for realistic effects such as multiport interactions and coupling.

In particular, many contributions in the literature rely on \emph{cascade-based} models with unilateral inter-layer propagation. Although convenient, these assumptions are often difficult to justify electromagnetically and may conflict with reciprocity in compact passive implementations. Building on existing multiport formulations, this paper further develops an optimization-oriented S-parameter model that includes additional non-ideal effects (e.g., tunable-circuit and return losses) and validates both models and optimization algorithms through full-wave simulations.

To bridge signal-processing-oriented and electromagnetics-oriented perspectives, we employ FEKO\textsuperscript{\textregistered}\footnote{https://altairengineering.fr/feko/.} to validate an end-to-end S-parameter channel representation, optimize the SIM tuning vector $\boldsymbol{\eta}$, and then re-inject the optimized circuit configuration into electromagnetic simulations. This verification-driven workflow aims at providing more credible and practically grounded SIM design guidelines by systematically connecting optimization-friendly abstractions with electromagnetic validation.

The remainder of this paper is organized as follows. Section~III introduces the CEO S-parameter model and the associated end-to-end input-output representation. Section~IV presents the S-parameter representation and optimization framework for a fully coupled non-isolated SIM. Section~V discusses simplified SIM models and optimization, including the layered isolated and weakly-coupled cases. Section~VI addresses optimization under discrete cell-state sets. Section~VII reports the electromagnetic simulation setup and numerical results for communication and sensing scenarios. Finally, Section~VIII concludes the paper.
\section{CEO S-parameter model}
\label{sec:linear_multiport}

\begin{figure}[t]
	\centering
	\includegraphics[width=1\columnwidth]{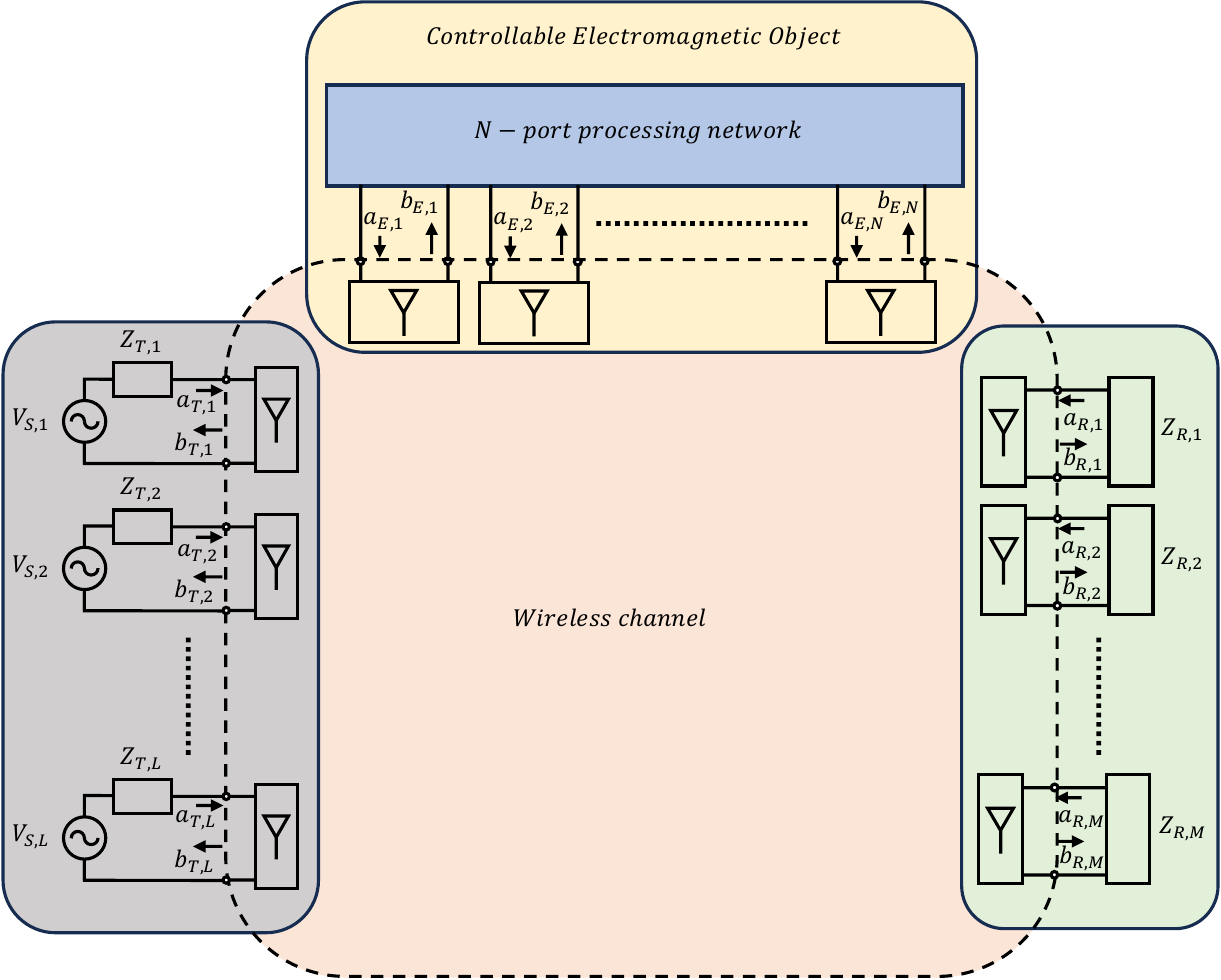}
	\caption{Multiport network representation of the considered CEO scenario, including a transmitting array (T), an $N$-port CEO, and a receiving array (R).}
	\label{fig:eco_mpn_model}
\end{figure}

This section briefly recalls the electromagnetic
scenario comprising transmitting and receiving antenna arrays and an intermediate controllable electromagnetic object (CEO) presented in \cite{abrardo_E}. To elaborate, we consider the multi-port network model introduced in \cite{abrardo_E} to characterize wave propagation among a transmitting active antenna (T) with $L$ ports, a CEO with $N$ ports, and a receiving antenna (R) with $M$ ports. The overall system is modeled as a linear $N_{t}=(L+N+M)$-port network, described by a scattering matrix
$\bm S \in \mathbb{C}^{N_t\times N_t}$, which relates incident and reflected waves as illustrated in Fig.~\ref{fig:eco_mpn_model}.
Using standard power-wave variables, the incident and reflected vectors satisfy
\begin{equation}
\bm b = \bm S \bm a,
\qquad
\bm a,\bm b\in\mathbb{C}^{N_t}.
\end{equation}
We partition the ports into three groups
\begin{equation}
\bm a=\begin{bmatrix}\bm a_T \\ \bm a_E \\ \bm a_R\end{bmatrix},\qquad
\bm b=\begin{bmatrix}\bm b_T \\ \bm b_E \\ \bm b_R\end{bmatrix},\qquad
\bm S=\begin{bmatrix}
\bm S_{TT} & \bm S_{TE} & \bm S_{TR}\\
\bm S_{ET} & \bm S_{EE} & \bm S_{ER}\\
\bm S_{RT} & \bm S_{RE} & \bm S_{RR}
\end{bmatrix}.
\end{equation}
As in \cite{abrardo_E}, we assume matched transmitter/receiver ports, i.e.,
\[
\bm a_T=\bm a_S,\qquad \bm a_R=\bm 0.
\]
The internal and output relations then read
\begin{align}
\bm b_E &= \bm S_{ET}\bm a_S + \bm S_{EE}\bm a_E,
\label{eq:bE_general}\\
\bm y \triangleq \bm b_R &= \bm S_{RT}\bm a_S + \bm S_{RE}\bm a_E.
\label{eq:y_general_vec}
\end{align}
The CEO termination is modeled through a (generally reconfigurable) $N$-port network:
\begin{equation}
\bm a_E = \bm\Gamma(\boldsymbol{\eta})\,\bm b_E,
\label{eq:Gamma_global}
\end{equation}
where the termination matrix $\bm\Gamma(\boldsymbol{\eta})\in\mathbb{C}^{N\times N}$ depends on a set of control parameters $\boldsymbol{\eta}$.

By direct algebraic manipulation (see~\cite{abrardo_E}), the end-to-end input--output relation is
%
\begin{equation}
\bm y = \big(\bm S_{RT} +\bm S_{RE}\,\bm T(\boldsymbol{\eta})\,\bm S_{ET}\big)\bm a_S.
\label{eq:GTF_def_with_direct}
\end{equation}
where $\bm T(\boldsymbol{\eta}) = (\bm\Gamma^{-1}(\boldsymbol{\eta})-\bm S_{EE})^{-1}$.
\section{S-parameters representation and optimization of a SIM}
\label{sec:SIM_model}

\begin{figure}[t]
	\centering
	\includegraphics[width=1\columnwidth]{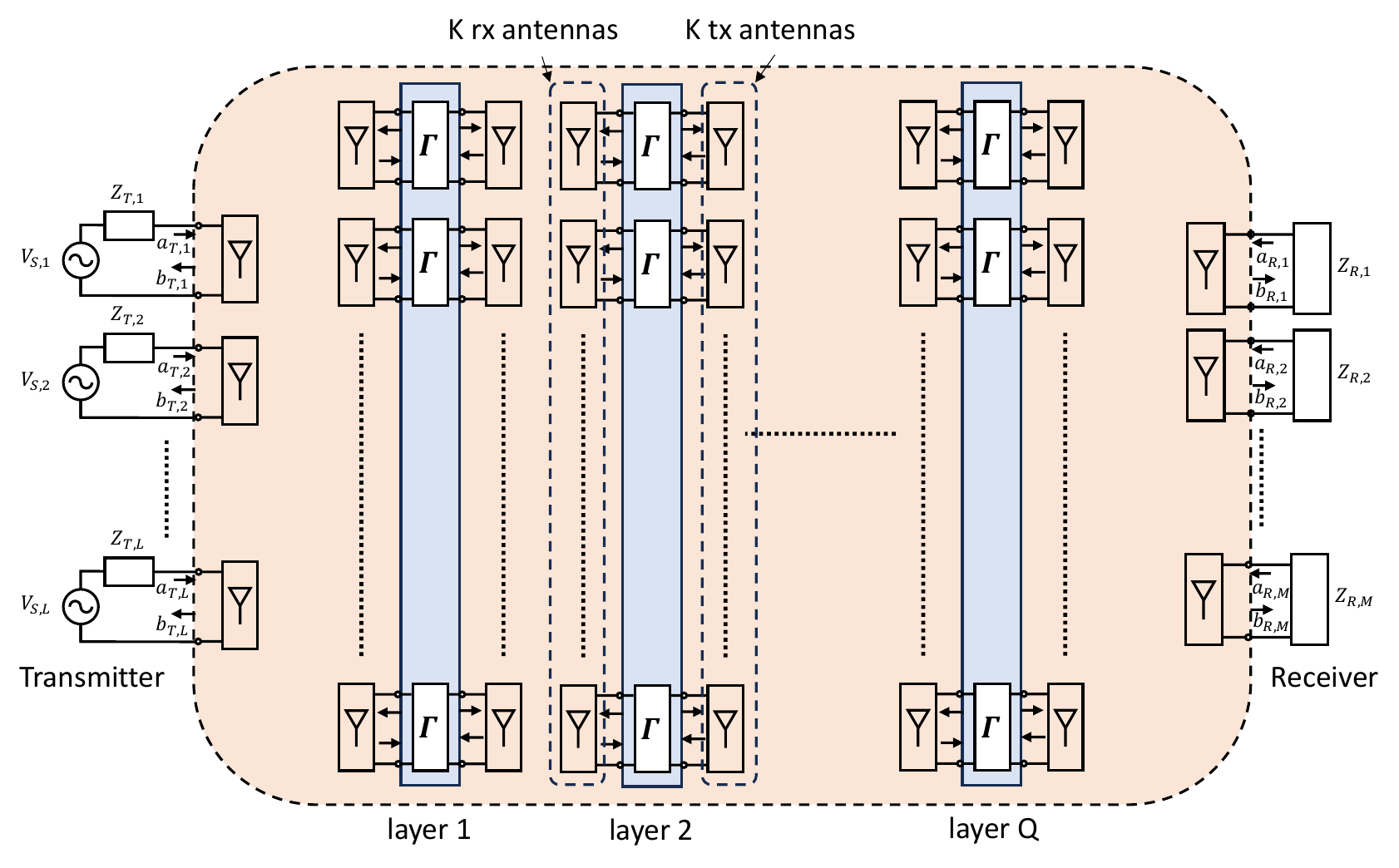}
	\caption{\textcolor{black}{Schematic of the considered SIM as a CEO with a layered architecture. The shaded region represents the electromagnetic interactions among all antenna ports due to wave propagation in the surrounding medium.}}
	\vspace{-2mm}
	\label{fig:sim_schematic}
\end{figure}

\begin{figure}[t]
	\centering
	\includegraphics[width=1\columnwidth]{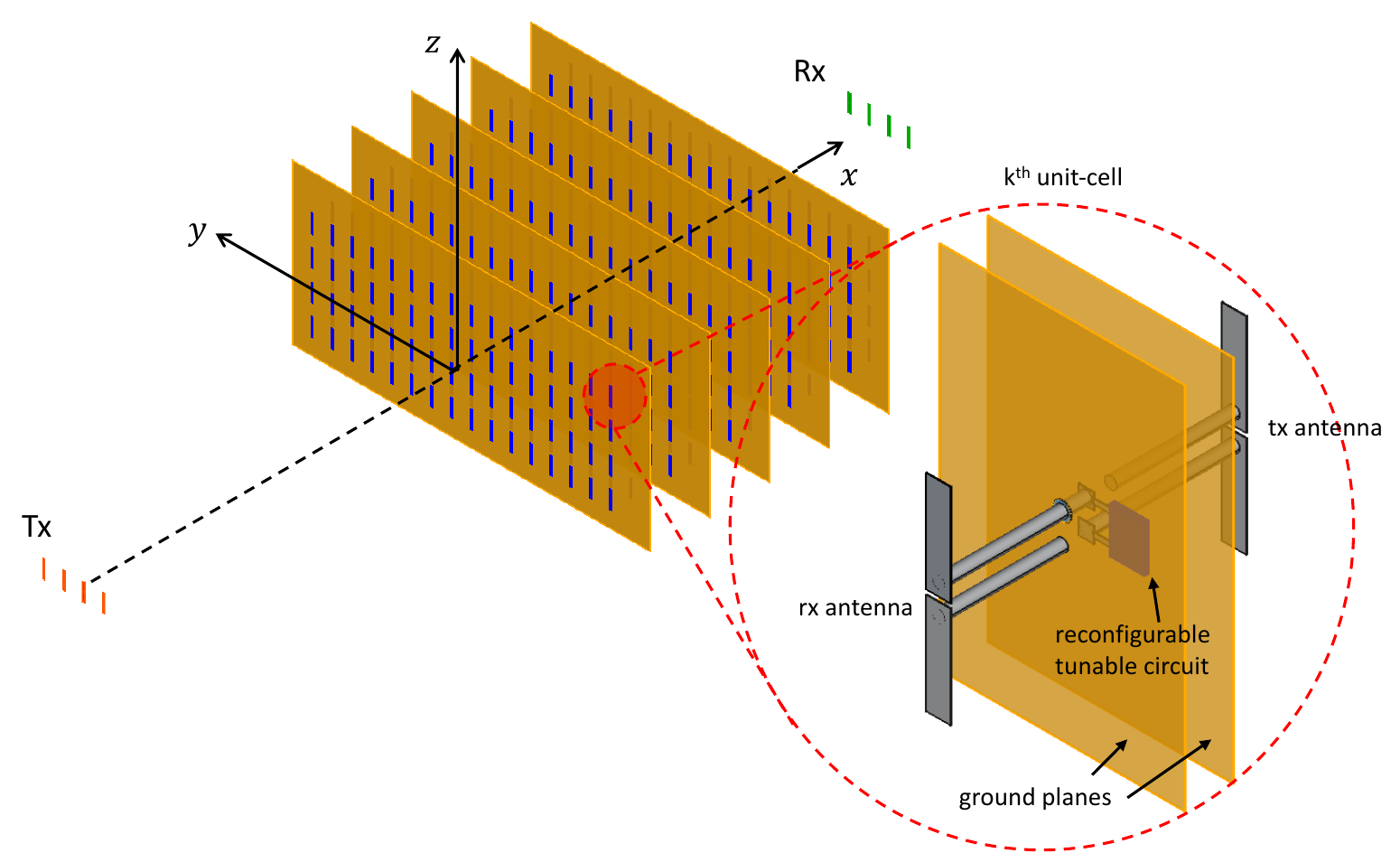}
	\caption{\textcolor{black}{SIM geometry and unit-cell example}}
	\vspace{-2mm}
	\label{fig:sim_schematic_3D}
\end{figure}

We consider an SIM implemented as a CEO composed of $Q$ transmissive
reconfigurable intelligent surfaces (T-RISs). 


In the following, each
T-RIS is referred to as a \emph{layer} of the SIM.
Each layer has the typical structure of a transmittarray and consists of
two antenna arrays separated by one or more ground planes: a \emph{receive array}
that collects the incident electromagnetic field, and a \emph{transmit array}
that re-radiates a transformed field toward the opposite side of the structure.
Both the receive and transmit arrays contain $K$ antennas, which are
modeled as $K$ ports on each array. Consequently, a single layer contributes
$2K$ antennas to the SIM network. Since the SIM is composed of $Q$ layers,
the total number of antenna ports is 
\begin{equation}
	N = 2QK .
	\label{eq:N_ports_SIM}
\end{equation}
The two arrays are interconnected through tunable networks implementing a
programmable transformation, typically a phase shift. In this work we adopt a
\emph{block-diagonal} interconnection architecture, where each receive antenna of a layer
is connected only to its corresponding layer transmit antenna through an internal, independent
two-port network. Therefore, each layer is composed of $K$ independent tunable
two-port cells, each coupling one receive antenna to one transmit antenna, as
illustrated in Fig.~\ref{fig:sim_schematic}.

Figure~\ref{fig:sim_schematic_3D} shows an example of a SIM structure in which
simple strip dipoles are used as antennas. The structure is composed of five
layers, illuminated by four transmitting antennas and sensed by four receiving
antennas.
To better illustrate the possible configuration of a single $k$-th unit cell,
the inset of Fig.~\ref{fig:sim_schematic_3D} reports an example of a
transmittarray unit cell. The receiving and transmitting antennas are realized
as half-wavelength resonant strip dipoles suspended above their respective
ground planes. 
The two ground planes can be electrically connected through metallic
vias (not explicitly shown in the inset) so that they are maintained at
the same ground potential. The region between the two ground
planes can host a reconfigurable phase-shifting device, such as a MMIC phase
shifter, or alternatively a switching network implementing different stripline
transmission-line lengths connected to the strip dipoles through simple
conducting wires. The same region may also accommodate the required control
circuitry. It is important to emphasize that this unit-cell configuration is provided only
as a representative example of a receive–transmit structure interconnected by a
tunable network, as commonly adopted in transmitarray architectures
\cite{2017Abdelrahman}.

From a network-theoretic perspective, the electromagnetic coupling among all the
antenna ports, arising from wave propagation in the surrounding medium (e.g.,
free space), is modeled as a linear multiport scattering network described by the matrix
$ \bm S_{EE}\in\mathbb{C}^{N\times N}$

The SIM antenna ports are globally indexed as
$ \{1,\ldots,N\},$with $N=2QK$.
Using a layer-wise ordering, the $q$-th layer ports occupy the index range
$ \{\,2K(q-1)+1,\ldots,2Kq\,\}. $
Within a given layer, the $k$-th cell couples the two ports with
local indices $k$ and $k+K$. Accordingly, the global indices of the
two ports associated with the $k$-th cell of the $q$-th layer are
\begin{subequations}\label{eq:cell_indices_rigorous}
	\begin{align}
		m(q,k) &= 2K(q-1)+k, \\
		n(q,k) &= 2K(q-1)+k+K,
	\end{align}
\end{subequations}
with $k\in\{1,\ldots,K\}$ and $q\in\{1,\ldots,Q\}$.

Since each of the $Q$ layers contains $K$ independent tunable two-port
networks, the SIM includes a total of $QK$ tunable cells. For notational
convenience, the cells are indexed by a single integer
\begin{equation}
	p\in\{1,\ldots,QK\}.
	\label{eq:ps_index}
\end{equation}
For a given layer $q$ and cell $k$, the corresponding index is
$p=K(q-1)+k$. The two ports associated with this cell are denoted by
$m(p)$ and $n(p)$, where $m(p)=m(q,k)$ and $n(p)=n(q,k)$.
For the cell controlled by the parameter $\eta_p$, the scattering
relation between the corresponding antenna ports is
\begin{equation}
	\begin{bmatrix}
		a_{E,m(p)}\\[2pt]
		a_{E,n(p)}
	\end{bmatrix}
	=
	\bm \Gamma_{\mathrm{cell}}\!\left(\eta_p\right)
	\begin{bmatrix}
		b_{E,m(p)}\\[2pt]
		b_{E,n(p)}
	\end{bmatrix},
	\label{eq:cell_scattering_rigorous}
\end{equation}
where $\bm \Gamma_{\mathrm{cell}}(\eta_p)\in\mathbb{C}^{2\times2}$ is the
scattering matrix of the tunable cell, which may in general be lossy
and mismatched.
The widely used \emph{ideal phase-shifter} model is obtained as a
special case and is reported here only as a convenient reference:
\begin{equation}
	\bm \Gamma_{\mathrm{cell}}(\eta_p)
	=
	\begin{bmatrix}
		0 & e^{j\eta_p}\\
		e^{j\eta_p} & 0
	\end{bmatrix}.
	\label{eq:Scell_rigorous}
\end{equation}
In the general case, $\eta_p$ controls the transmission properties of
the cell, namely the coefficients $\Gamma_{2,1}(\eta_p)$ and
$\Gamma_{1,2}(\eta_p)$, while the diagonal entries
$\Gamma_{1,1}(\eta_p)$ and $\Gamma_{2,2}(\eta_p)$ may be nonzero due to
return loss and other non-idealities. Likewise, the magnitudes
$|\Gamma_{2,1}(\eta_p)|$ and $|\Gamma_{1,2}(\eta_p)|$ can be smaller than
one, accounting for insertion loss, as discussed later when practical
circuit realizations are introduced.

Collecting all the $QK$ tunable parameters into
$\boldsymbol{\eta}=[\eta_1,\ldots,\eta_{QK}]^T$, the SIM terminations are
described by the global scattering relation in \eqref{eq:Gamma_global}.
The matrix $\bm\Gamma(\boldsymbol{\eta})\in\mathbb{C}^{N\times N}$ is
block-diagonal and composed of $Q$ non-overlapping $2K\times2K$ blocks
(one per layer). Each parameter $\eta_p$ affects only one $2\times2$
submatrix associated with the port pair $(m(p),n(p))$.

Because of this structure, the inverse matrix
$\bm\Gamma^{-1}(\boldsymbol{\eta})$ has the same sparsity pattern and can
be obtained by inverting each $2\times2$ block
$\bm\Gamma_{\mathrm{cell}}(\eta_p)$ independently.

\color{black}

\subsection{SIM optimization problem}

The SIM configuration task can be interpreted as a regression problem for a
parametrized physics-based input--output model. Given a finite set of input
excitations and desired output responses, the SIM control parameters are
optimized by minimizing a suitable loss function.
Specifically, let $I$ input excitations be collected in the source matrix
\[
\bm A_S=\big[\bm a_S^{(1)},\ldots,\bm a_S^{(I)}\big],
\]
and define the corresponding internal and output matrices
\[
\bm A_E=\big[\bm a_E^{(1)},\ldots,\bm a_E^{(I)}\big], \qquad
\bm B_E=\big[\bm b_E^{(1)},\ldots,\bm b_E^{(I)}\big],
\]
\[
\bm Y=\big[\bm y^{(1)},\ldots,\bm y^{(I)}\big].
\]
Accordingly,
\begin{align}
\bm B_E &= \bm S_{ET}\bm A_S + \bm S_{EE}\bm A_E,
\label{eq:BE_mat}\\
\bm Y   &= \bm S_{RT}\bm A_S + \bm S_{RE}\bm A_E.
\label{eq:Y_mat}
\end{align}
Given a desired output matrix $\bm Y_d\in\mathbb{C}^{M\times I}$, the SIM
configuration is obtained by minimizing a loss function that measures the
mismatch between $\bm Y$ and $\bm Y_d$. Let $\beta\in\mathbb{C}$ denote a
complex scaling factor and define the error matrix as
\[
\bm E \triangleq \beta\,\bm Y - \bm Y_d .
\]
The loss function is then defined as
\begin{equation}
L=\|\bm E\|_F^2,
\label{eq:loss}
\end{equation}
where $\|\cdot\|_F$ denotes the Frobenius norm.
The SIM optimization problem therefore consists of determining the vector of
control parameters $\boldsymbol{\eta}$ that minimizes the loss function in
\eqref{eq:loss}. In this work, the minimization is performed through a
gradient-based iterative procedure combined with a backtracking line-search
satisfying the Armijo condition. This choice provides a robust and widely used
strategy for the optimization of programmable electromagnetic structures.
Accordingly, once the gradient of the loss function with respect to the SIM
parameters $\boldsymbol{\eta}$ is available, the configuration vector can be
updated through a standard descent iteration. For this reason, in the
remainder of the paper we focus on deriving efficient expressions for the
gradient of the loss function under the different SIM models considered.
In practice, providing the gradient expressions for each model directly
defines the corresponding optimization procedure used to configure the SIM.

\subsection{Gradient evaluation}
\label{sec:Gradient evaluation}

Combining \eqref{eq:BE_mat} with \eqref{eq:Gamma_global} yields the fixed-point system of equations
\begin{equation}
(\bm I-\bm\Gamma\bm S_{EE})\bm A_E
=
\bm\Gamma\bm S_{ET}\bm A_S,
\label{eq:AE}
\end{equation}
and
\begin{equation}
(\bm I-\bm S_{EE}\bm\Gamma)\bm B_E
=
\bm S_{ET}\bm A_S.
\label{eq:BE}
\end{equation}
Differentiating \eqref{eq:AE} with respect to the phase parameter $\eta_p$ associated with the $p$-th unit cell yields
\begin{equation}
(\bm I-\bm\Gamma\bm S_{EE})
\frac{\partial\bm A_E}{\partial\eta_p}
=
\frac{\partial\bm\Gamma}{\partial\eta_p}\,\bm B_E.
\label{eq:dAE_linear}
\end{equation}
Let $\bm E=\beta\bm Y-\bm Y_d$ and recall that
$L=\|\bm E\|_F^2=\langle\bm E,\bm E\rangle$, where
$\langle \bm X,\bm Y\rangle \triangleq \mathrm{trace}(\bm X^H\bm Y)$.
Using the relation $\delta\bm Y=\bm S_{RE}\delta\bm A_E$, the first-order variation of the loss function is
\begin{equation}
\delta L
=
2\,\Re\!\left\langle
\bm S_{RE}^H(\beta^*\bm E),
\delta\bm A_E
\right\rangle.
\label{eq:deltaL_linear}
\end{equation}
We define the adjoint forcing vector as
\begin{equation}
\bm q
\triangleq
\bm S_{RE}^H(\beta^*\bm E),
\label{eq:q_linear}
\end{equation}
and introduce the adjoint variable $\bm U\in\mathbb{C}^{N\times I}$ as the
solution of the adjoint system
\begin{equation}
(\bm I-\bm\Gamma\bm S_{EE})^H\bm U
=
\bm q.
\label{eq:adj_linear}
\end{equation}
Then, the gradient of the loss function with respect to $\eta_p$ can be written as
\begin{equation}
\frac{\partial L}{\partial\eta_p}
=
2\,\Re\!\left\langle
\bm U,
\frac{\partial\bm\Gamma}{\partial\eta_p}\bm B_E
\right\rangle.
\label{eq:grad_linear}
\end{equation}
Due to the assumption of diagonal T-RISs, the derivative of the termination matrix
$\bm\Gamma$ with respect to $\eta_p$ can be computable provided that an analytical model for the cell matrix $\bm \Gamma_{\mathrm{cell}}(\eta_p)$ is available.
Specifically,
\begin{equation}
\frac{\partial\bm\Gamma}{\partial\eta_p}
=
\bm T_p,
\label{eq:Tangent_matrix_def}
\end{equation}
where $\bm T_p\in\mathbb{C}^{N\times N}$ inherits the sparsity induced by the diagonal two-port architecture: it has at most four non-zero entries, corresponding to the $2\times2$ block associated with the port pair $(m(p),\,n(p))$.

The computational cost is dominated by the solution of the linear system of equations \eqref{eq:BE} and \eqref{eq:adj_linear}, which needs to be solved once per gradient evaluation. 
Consequently, the overall complexity scales cubically with the
number of antennas, as follows:
\begin{equation}
\mathcal{C}_{{SIM-NI}}
=
\mathcal{O}\!\left(N^3\right),
\label{eq:complexity_gSIM}
\end{equation}
where $SIM-NI$ refers to a non-isolated (fully coupled) SIM configuration.
$\mathcal{C}_{{SIM-NI}}$is obtained without exploiting any specific structural properties of the SIM, such as its layered architecture, possible sparsity patterns, or assumptions of electromagnetic isolation between layers
or between the SIM internal layers and the transmitting/receiving arrays.
Although idealized, such assumptions can be regarded as approximately valid for specific SIM realizations and design choices. As will be shown in the following, explicitly accounting for the layered structure of the SIM enables significant reductions in computational complexity.
\section{Simplified SIM models and optimization}
\subsection{S-parameters representation of a layered isolated SIM}
\label{sec:ideal_SIM_model}

In this section we introduce a simplified SIM model obtained by reformulating
in the S-parameter domain the impedance-based model proposed in~\cite{Abra_SIM1}.
The physical assumptions adopted here are identical to those used in that work,
and therefore only the essential elements are briefly recalled.

\vspace{-1.5mm}
\begin{assumption}[Isolation from transmitting/receiving arrays]\label{ass:A1}
The SIM is deployed such that the electromagnetic coupling between the SIM
internal layers and the transmitting/receiving arrays is negligible, except
for the intended illumination and observation links captured by the
multiport network model.
\end{assumption}

\vspace{-1.5mm}
\begin{assumption}[Layer coupling structure]\label{ass:A2}
Within each T-RIS, the two arrays composing the transmissive cell are not
directly coupled except through the tunable circuit network. Across the SIM,
electromagnetic coupling between different layers occurs only through
propagation in the surrounding medium between adjacent arrays.
\end{assumption}
\vspace{-1mm}

Although idealized, these assumptions are physically meaningful and can be
reasonably approached in practice through suitable design strategies, such as
the introduction of ground planes between adjacent T-RIS layers and the use of
absorbing materials to suppress unwanted electromagnetic leakage.

\begin{figure}[t]
  \centering
  \includegraphics[width=1\columnwidth]{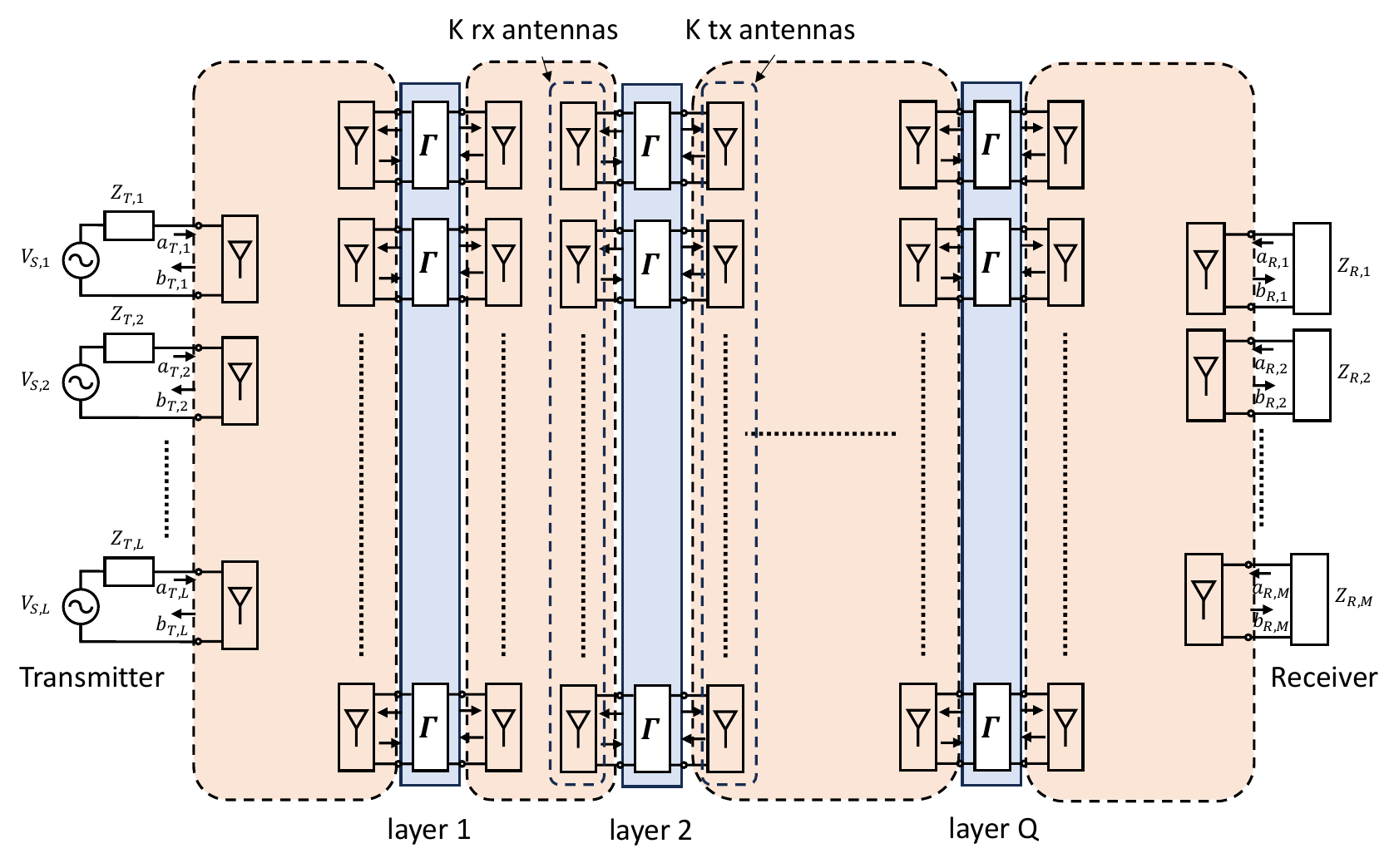}
  \caption{\textcolor{black}{Illustration of the layer-isolated SIM ($SIM$-$I$) coupling structure. 
The shaded regions represent electromagnetic coupling only between adjacent 
arrays: from the transmitter to the receive array of the first layer, between 
the transmit and receive arrays of consecutive layers, and from the transmit 
array of the last layer to the receiver. Interactions between non-adjacent 
layers are neglected}}
  \vspace{-2mm}
  \label{fig:simI_schematic}
\end{figure}

Under Assumptions~\ref{ass:A1}--\ref{ass:A2}, the SIM exhibits a structured
electromagnetic coupling pattern in which only neighboring layers interact
directly. As illustrated in Fig.~\ref{fig:simI_schematic}, the transmitter
couples only to the first SIM layer, the receiver only to the last layer, and
interactions inside the SIM occur only between adjacent layers.

These assumptions induce a structured sparsity in the internal scattering
matrix $\bm S_{EE}$, which becomes block diagonal (or block-banded with
nearest-neighbor interactions) when the ports are ordered according to the SIM
layers. This structure is fully analogous to the impedance-matrix structure
obtained in~\cite{Abra_SIM1}. Consequently, the same algebraic framework and
computational strategy derived in~\cite{Abra_SIM1} can be directly applied in
the S-parameter formulation.

Importantly, this model does \emph{not} correspond to a cascade representation
of the SIM. In the proposed multiport formulation all ports remain mutually
coupled through the global scattering matrix, and the tunable circuits
connecting the ports are intrinsically bidirectional. As a result,
electromagnetic propagation inside the structure is never forced to be
unilateral and reciprocity is naturally preserved. This distinguishes the
present formulation from cascade-based abstractions often adopted in the SIM
literature.

For these reasons, the model will be referred to as the \emph{layered isolated
SIM} (SIM-I).

\subsection{Gradient evaluation for a layered isolated SIM}
\label{sec:grad_ideal_SIM}

The gradient computation follows the general framework introduced in
Section~\ref{sec:linear_multiport}. Due to the structured sparsity induced by
Assumptions~\ref{ass:A1}--\ref{ass:A2}, the matrix
\[
\bm\Gamma^{-1}-\bm S_{EE}
\]
exhibits the same block-banded structure identified in the impedance-domain
formulation of~\cite{Abra_SIM1}. Consequently, the recursive algorithms
derived in~\cite[Appendices~A--B]{Abra_SIM1} can be directly employed to
compute the quantities required for the gradient evaluation.

As a result, the computational complexity of the gradient evaluation scales as
\begin{equation}
\mathcal{C}_{SIM-I}
=
\mathcal{O}\!\left(QK^3\right),
\end{equation}
that is, cubic in the number of ports per array and linear in the number of
SIM layers, exactly as in the impedance-based formulation of~\cite{Abra_SIM1}.

\subsection{Weakly-coupled SIM}
\label{subsec:SIMQ}

We now introduce an intermediate SIM model, referred to as the \emph{weakly-coupled SIM (SIM-W)} model, which relaxes Assumptions~\ref{ass:A1} and~\ref{ass:A2}.
As a consequence, additional electromagnetic coupling interactions are admitted with respect to the isolated SIM model, although the coupling is assumed to be weak.

In particular, we focus on the internal scattering matrix $\bm S_{EE}$.
Instead of enforcing a perfectly block-diagonal structure, we model $\bm S_{EE}$ as the superposition of a dominant block-structured component and a residual perturbation, namely
\begin{equation}
\bm S_{EE}
=
\bm S_{EE}^{(0)} + \Delta \bm S,
\label{eq:SEE_SIMQ}
\end{equation}
where $\bm S_{EE}^{(0)}$ has the block-diagonal structure as the \emph{SIM-I} case, while $\Delta \bm S$ collects all residual inter-block coupling terms neglected in the \emph{SIM-I} formulation.
By construction, the additional coupling is assumed to be weak, so that
\begin{equation}
\|\Delta \bm S\| \ll \|\bm S_{EE}^{(0)}\|.
\end{equation}
Under this assumption, the SIM response depends on the inverse matrix
\begin{equation}
\left(\bm\Gamma^{-1}-\bm S_{EE}\right)^{-1}
=
\left(\bm\Gamma^{-1}-\bm S_{EE}^{(0)}-\Delta \bm S\right)^{-1}.
\label{eq:core_inverse_SIMQ}
\end{equation}
Defining
\begin{equation}
\bm A \triangleq \bm\Gamma^{-1}-\bm S_{EE}^{(0)},
\end{equation}
and assuming that the perturbation satisfies
\begin{equation}
\left\|\bm A^{-1}\Delta \bm S\right\| \ll 1,
\label{eq:Neumann_condition_SIMQ}
\end{equation}
the inverse in \eqref{eq:core_inverse_SIMQ} can be approximated by means of
the Neumann series as
\begin{align}\label{eq:Neumann_approx_SIMQ}
\left(\bm A-\Delta \bm S\right)^{-1}
&=
\bm A^{-1}
\sum_{n=0}^{\infty}
\left(\bm A^{-1}\Delta \bm S\right)^n 
\\
&\approx
\bm A^{-1}
+
\bm A^{-1}\Delta \bm S\,\bm A^{-1},
\nonumber 
\end{align}
where only the first-order correction term is retained.

From a physical standpoint, the SIM-W model captures practical SIM architectures in which the assumption of electromagnetic isolation is legitimate only approximately.
For instance, ground planes inserted between adjacent T-RIS layers, absorbing materials, or sufficient inter-layer spacing can strongly attenuate undesired couplings without completely eliminating them.
In such scenarios, residual inter-block interactions are unavoidable but may remain sufficiently weak to be treated as small perturbations.

The SIM-W model therefore strikes a balance between physical realism and computational tractability.
On the one hand, it relaxes the strong assumptions underlying the ideal SIM formulation; on the other hand, it preserves the block-banded structure of the dominant term $\bm S_{EE}^{(0)}$, which can be exploited to significantly reduce the computational burden with respect to a fully general multiport model.

The computational complexity of the SIM-W model is
dominated by the evaluation of the inverse matrix
\(
(\bm\Gamma^{-1}-\bm S_{EE})^{-1}
\),
which is approximated according to \eqref{eq:Neumann_approx_SIMQ}.
The leading contribution corresponds to the computation of
\(
\bm A^{-1}=(\bm\Gamma^{-1}-\bm S_{EE}^{(0)})^{-1}
\),
while the first-order correction requires additional multiplications by $\Delta \bm S$.

Since $\bm S_{EE}^{(0)}$ has the same block-banded structure as in the ideal SIM case, the matrix $\bm A$ can be inverted using block-recursive techniques similar to those developed in \cite{Abra_SIM1}.
However, unlike the isolated SIM formulation---where only the first $K$ columns and the last $K$ rows of $\bm A^{-1}$ are required---the SIM-W model
necessitates the computation of the full inverse matrix in order to account
for residual inter-block couplings.

As a result, the overall computational complexity of the SIM-W model scales
as
\begin{equation}
\mathcal C_{{SIM-W}}
=
\mathcal O\!\left(Q^2K^3\right),
\end{equation}
where $Q$ denotes the number of T-RIS layers and $K$ the number of ports per layer.
Despite this increase in complexity with respect to the isolated SIM case, the resulting complexity remains significantly lower than that of a non-isolated SIM
model, which requires the inversion of a dense $(2QK)\times(2QK)$ matrix and
scales as $\mathcal O\!\left((2QK)^3\right)$.

The detailed block-recursive algorithm used to compute the full inverse
matrix $\bm A^{-1}$ is reported in Appendix~\ref{app:full_T}.
\section{Optimization with discrete cell-state sets}
\label{subsec:discrete_optimization}

In practical SIM implementations, the tunable elements are constrained to a finite set of admissible electromagnetic configurations dictated by the underlying hardware and circuit realizations. 
To explicitly account for such constraints, we consider an optimization problem when each $2\times2$ scattering cell can assume one of $|\mathcal P|$ \emph{discrete states}.

\subsection{Discrete scattering-cell model}

Let $p\in\{1,\ldots,QK\}$ index all the tunable cells of a diagonal T-RIS SIM. The
$p$-th cell is characterized by a finite codebook of admissible $2\times2$
scattering matrices
\begin{align}\label{eq:Scell_discrete_general}
\bm \Gamma_{\mathrm{cell,D}}^{(p)}(\ell)
= &
\begin{bmatrix}
\Gamma_{\mathrm{cell,D}}^{(p)}(\ell)_{1,1} & \Gamma_{\mathrm{cell,D}}^{(p)}(\ell)_{1,2}\\
\Gamma_{\mathrm{cell,D}}^{(p)}(\ell)_{2,1} & \Gamma_{\mathrm{cell,D}}^{(p)}(\ell)_{2,2}
\end{bmatrix},
\\
& \text{with }
\ell\in\mathcal P\triangleq\{1,\ldots,|\mathcal P|\}.
\nonumber
\end{align}
Each discrete state may capture non-ideal effects such as insertion losses, residual reflections, or amplitude variations, while satisfying the physical constraints imposed by the considered hardware (e.g., passivity and reciprocity,
when applicable).

The global termination matrix $\bm\Gamma$ is block diagonal with respect to the cell partition, and its $2\times2$ $p$-th block is selected from the codebook in  \eqref{eq:Scell_discrete_general}. 
Let $\boldsymbol{\ell}=[\ell_1,\ldots,\ell_{QK}]^T$ denote the vector collecting the discrete states of all cells.

\paragraph{Special case (phase-only quantization)}
A widely adopted idealized model corresponds to a phase-only transmissive cell In this case, the $2\times2$ network model is
\begin{equation}
\bm \Gamma_{\mathrm{cell,D}}(\phi)
=
\begin{bmatrix}
0 & e^{j\phi}\\
e^{j\phi} & 0
\end{bmatrix},
\label{eq:Scell_phase_only}
\end{equation}
where $\phi$ is a continuous phase. If the hardware supports a finite set of
phases $\Phi=\{\phi^{(1)},\ldots,\phi^{(|\Phi|)}\}$, then
\eqref{eq:Scell_discrete_general} is recovered by setting
$\bm \Gamma_{\mathrm{cell,D}}^{(p)}(\ell)=\bm \Gamma_{\mathrm{cell,D}}(\phi^{(\ell)})$ and
$\mathcal P=\{1,\ldots,|\Phi|\}$.

\subsection{Initialization via projection onto the codebook}

A parametric approach to solve the discrete optimization problem is to initialize it assuming continuous phase-only
cells. Let $\bm \Gamma_{\mathrm{cell,D},p}^\star$ denote the $2\times2$ scattering matrix corresponding to the continuous solution at cell $p$. A discrete-feasible initialization is obtained by projecting each cell onto its codebook via
\begin{equation}
\ell_p^{(0)}
=
\arg\min_{\ell\in\mathcal P}
\left\|
\bm \Gamma_{\mathrm{cell,D}}^{(p)}(\ell) - \bm \Gamma_{\mathrm{cell,D},p}^\star
\right\|_F.
\label{eq:codebook_projection}
\end{equation}
In the phase-only case, \eqref{eq:codebook_projection} reduces to standard phase
quantization.

\subsubsection{Coordinate-wise discrete descent}

Starting from $\boldsymbol{\ell}^{(0)}$, a coordinate-wise discrete descent can then be adopted. 
At each step, a single cell index $p$ is selected while all remaining cells are kept fixed. The optimal discrete state for the $p$-th cell is obtained
by exhaustive search over the finite set $\mathcal P$:
\begin{equation}
\ell_p \leftarrow
\arg\min_{\ell\in\mathcal P}
L\!\left(\ell,\boldsymbol{\ell}_{-p}\right),
\label{eq:coord_update_discrete}
\end{equation}
where $\boldsymbol{\ell}_{-p}$ denotes the collection of all states except
$\ell_p$. One sweep over $p=1,\ldots,QK$ defines an outer iteration.

By construction, the loss function is non-increasing at each update, and since the search space is finite, the algorithm converges in a finite number of steps to, in general, a local optimum.

\subsubsection{Fast loss evaluation via low-rank updates}

A naive evaluation of \eqref{eq:coord_update_discrete} would require recomputing,
for each candidate $\ell\in\mathcal P$, the inverse matrix associated with the SIM
response, typically of the form
\begin{equation}
\bm X(\boldsymbol{\ell})
=
\left(\bm\Gamma(\boldsymbol{\ell})^{-1}-\bm S_{EE}\right)^{-1},
\label{eq:X_def}
\end{equation}
which would be computationally prohibitive.

The key observation is that updating only the $p$-th cell state modifies
$\bm\Gamma^{-1}$ exclusively on the corresponding $2\times2$ block. Let
$\bm E_p\in\mathbb{R}^{N\times2}$ be the selector matrix extracting the two global
port indices associated with cell $p$. Then, for fixed
$\boldsymbol{\ell}_{-p}$,
\begin{equation}
\bm\Gamma(\ell_p,\boldsymbol{\ell}_{-p})^{-1}
=
\bm\Gamma(\boldsymbol{\ell}_{-p})^{-1}
+
\bm E_p\,\bm C_p(\ell_p)\,\bm E_p^T,
\label{eq:Gamma_inv_rank2}
\end{equation}
where $\bm C_p(\ell_p)\in\mathbb{C}^{2\times2}$ captures the difference between the
candidate inverse-cell block and the current one.

Defining
\begin{equation}
\bm A \triangleq \bm\Gamma(\boldsymbol{\ell}_{-p})^{-1}-\bm S_{EE},
\end{equation}
the SIM response matrix under a candidate update admits the rank-2 form
\begin{equation}
\bm X(\ell_p,\boldsymbol{\ell}_{-p})
=
\left(\bm A + \bm E_p\,\bm C_p(\ell_p)\,\bm E_p^T\right)^{-1}.
\label{eq:X_rank2}
\end{equation}
Applying the Woodbury identity yields
\begin{align}\label{eq:woodbury_rank2}
\bm X(\ell_p,\boldsymbol{\ell}_{-p})
=&
\bm A^{-1}
-
\bm A^{-1}\bm E_p
\left(\bm C_p(\ell_p)^{-1}
+ \right.\\
& \left.+
\bm E_p^T\bm A^{-1}\bm E_p\right)^{-1}
\bm E_p^T\bm A^{-1}.
\nonumber
\end{align}
Crucially, \eqref{eq:woodbury_rank2} requires only the two columns of
$\bm A^{-1}$ corresponding to the $p$-th cell, together with the inversion of a
$2\times2$ matrix. Therefore, once these columns are available, the evaluation
of each candidate $\ell\in\mathcal P$ has linear complexity $\mathcal O(N)$.

\subsubsection{Complexity discussion}

The computational burden of the proposed discrete optimization is dominated by
the cost of obtaining the required portions of $\bm A^{-1}$ for each coordinate
$p$. In contrast, the exhaustive search over the discrete set $\mathcal P$
introduces only a marginal overhead.

Specifically, for a fixed $p$, the evaluation of all $|\mathcal P|$ candidates
via \eqref{eq:woodbury_rank2} scales as $\mathcal O(|\mathcal P|\,N)$, which is
negligible compared to the cost of computing (or factorizing) $\bm A$, which, in turn, depends on the adopted SIM model.
\section{Simulation Results}
\label{sec:results}

In this section, we present a set of representative results obtained by
considering different SIM based functions. Specifically, we address
both classical communication-oriented tasks (hereafter referred to as
\emph{COMM scenarios}) and estimation-oriented tasks (referred to as
\emph{EST scenarios}). The main objective is to demonstrate that the proposed
system models and the associated optimization strategies are well suited for
the design of SIMs while consistently accounting for electromagnetic
propagation effects. To this end, the performance of the optimized SIM
configurations is assessed by means of electromagnetic simulations,
which provide an accurate representation of the underlying physical system.

\subsection{General Simulation Scenario}
\label{subsec:simulation_setup}

\begin{figure}
	\centering
	\includegraphics[width=1\columnwidth]{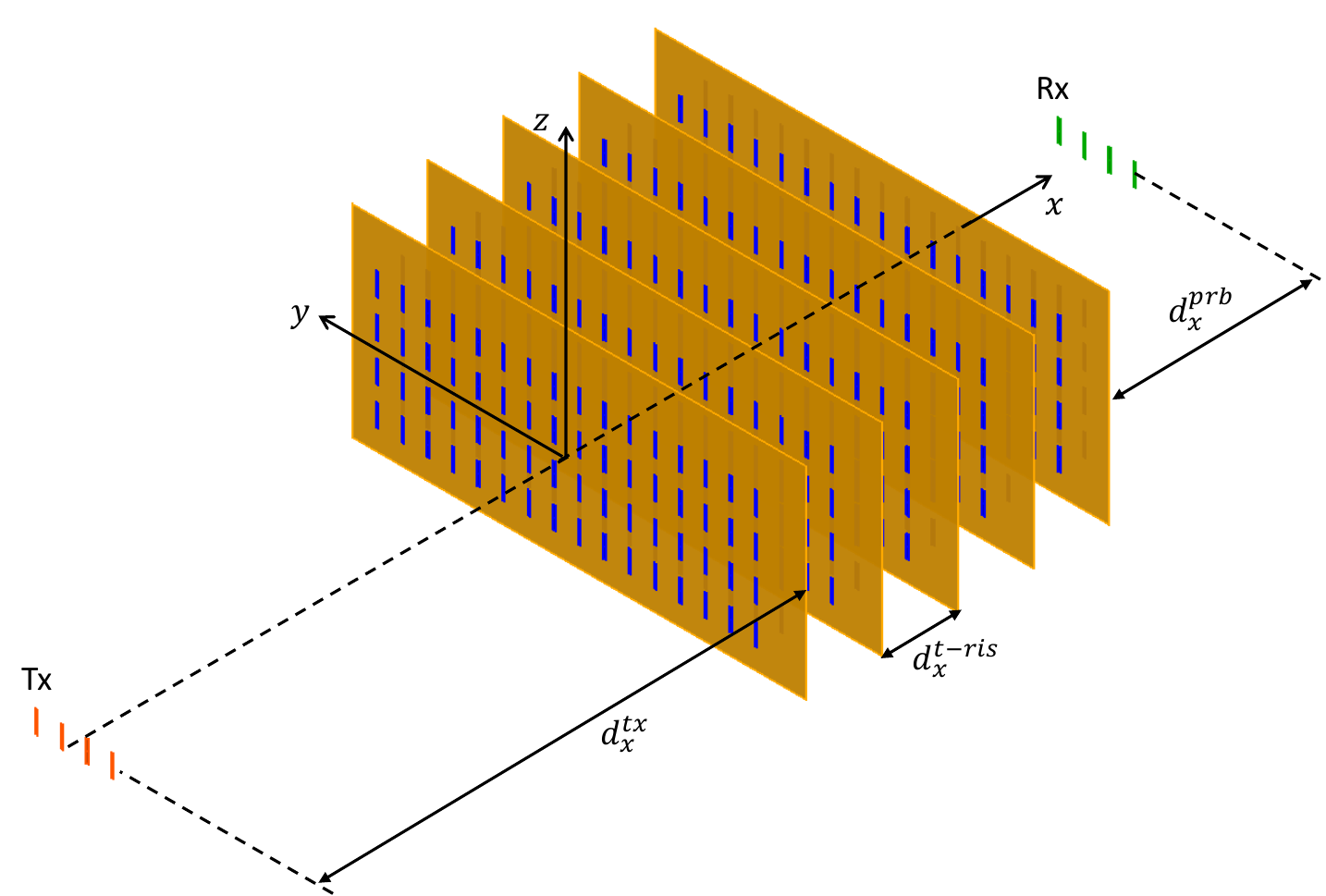}
	\caption{\textcolor{black}{Structure of the considered SIM with $Q$ stacked T-RIS layers. Each layer includes planar receive and transmit arrays interconnected by tunable phase-shifting networks.}}
	\vspace{-2mm}
	\label{fig:SIMmodel}
\end{figure}

The general structure of the SIM adopted in the following examples is shown
in Fig.~\ref{fig:SIMmodel}. It consists of $Q$ T-RIS layers arranged in a
stacked configuration, parallel and aligned along the $x$ axis of a global
reference system, and operating at a carrier frequency
$f_0 = 28$~GHz, corresponding to a free-space wavelength
$\lambda_0 = 10.7\ \mathrm{mm}$.

Each element of the $q$-th T-RIS, with $q = 1,2,\ldots,Q$, is modeled as a
center-fed planar metallic strip dipole of width $w_d = 0.05\lambda$
and length $l_d = 0.473\lambda$, oriented along the $y$ and $z$ axes,
respectively, and suspended over a ground plane at a distance
$h_d = 0.25\lambda$. The dipoles are arranged in a uniform planar array
(UPA) with $N_y^{q}$ elements along the $y$ axis and $N_z^{q}$ elements
along the $z$ axis. The inter-element spacing between adjacent dipoles is
$d_y^{dip} = \lambda/2$ along the $y$ direction and
$d_z^{dip} = (3/4)\lambda$ along the $z$ direction.

The spacing between two adjacent T-RIS layers, denoted as
$d_x^{\,t\text{-}ris}$, is identical for all $q$ and set to
$d_x^{\,t\text{-}ris} = 1.5\lambda$, resulting in a separation of
$\lambda$ between two facing antenna arrays. As sketched in
Fig.~\ref{fig:SIMmodel}, each T-RIS layer consists of a receive antenna
array and a transmit antenna array composed of $K$ unit cells. Each
unit cell consists of  one receiving and one transmitting strip
dipole located on opposite sides of the ground plane and internally
connected through a linear passive two-port network 
that enables tunability of the unit
cell, e.g., the applied phase shift.

A possible physical implementation of the unit cell based on strip
dipoles and tunable phase-shifting networks has been illustrated in
Fig.~\ref{fig:sim_schematic_3D} and is not repeated here for brevity.
In the present work, the detailed electromagnetic optimization of the
unit-cell layout is not addressed. Instead, to maintain a tractable
number of unknowns and enable full-wave simulations with finite arrays,
the modeling is limited to the strip dipoles with lumped ports located
at their center feeds. The interconnection between dipole ports is
introduced through circuit network files representing the phase shifters,
and the overall electromagnetic behavior is obtained through a
co-simulation between the full-wave solver and the SPICE circuit model
within the FEKO environment.

The overall electromagnetic behavior of the SIM is therefore determined
by the configuration of the $KQ$ tunable parameters collected in the vector
$\boldsymbol{\eta}$.

The SIM is excited at the first layer by a set of $N_t$ transmitting
antennas, which may be arranged in a UPA with $N_y^{tx}$ and
$N_z^{tx}$ elements along the $y$ and $z$ axes, respectively, or in a
generic configuration. Analogously, the processed electromagnetic field
is sensed by $N_R$ receiving antennas (probes), arranged in a UPA with
$N_y^{prb}$ and $N_z^{prb}$ elements along the $y$ and $z$ axes,
respectively, or in a generic configuration. Both the excitation and
probing antennas are modeled as strip dipoles identical to those
constituting the SIM, but radiating in free space.

In all examples, the matrices $\mathbf{S}_{EE}$ used in the optimization
process are obtained from the commercial full-wave simulator FEKO.

The adopted SIM topology has been selected with the twofold objective of
maintaining a reasonable computational effort when using 
MoM simulations while preserving a realistic and physically meaningful
structure in all considered scenarios. 
To this end, simulations with both finite and infinite ground
planes for the SIM layers have been carried out.

The adoption of an infinite ground plane ensures perfect isolation
between different SIM layers as well as between the SIM and the
transmit/receive ports, resulting in a band-diagonal S-parameter matrix
fully compatible with the $SIM\text{-}I$ model. Moreover, it provides a
significant reduction in computational time since a planar Green's
function is employed and the ground plane does not need to be
discretized into triangular elements.

When finite ground planes are employed, they are modeled as planar PEC
surfaces centered with respect to the UPA and having dimensions
$L_y^q=(N_y^q-1)d_y^{dip}+3\lambda$ and
$L_z^q=(N_z^q-1)d_z^{dip}+3\lambda$. In this case a more realistic but
computationally more demanding configuration is obtained, characterized
by generally nonzero mutual couplings and a fully populated S-parameter
matrix; accordingly, the $SIM\text{-}NI$ or $SIM\text{-}W$ models are
required.

In the finite ground-plane configuration, the SIM can be regarded as a
more realistic model since it inherently accounts for additional
electromagnetic couplings among different SIM layers and imperfect
isolation with respect to external sources. However, within the FEKO
simulation environment it was not feasible to include shielding structures or
absorbing panels that could effectively mitigate such couplings due to
their substantial modeling and computational complexity. As a result,
the finite-ground configuration represents an extreme non-isolated
scenario rather than a practically engineered SIM enclosure.

Moreover, no electromagnetic enclosure was implemented to confine a
significant fraction of the radiated power inside the SIM structure.
The simulated device therefore behaves as an open electromagnetic
system, inevitably radiating part of the power into the surrounding
free space and experiencing non-negligible power leakage.
This modeling limitation prevents the confinement of electromagnetic energy within the structure and directly impacts the interpretation of the results in terms of SNR. For this
reason, the reported SNR values should be understood as normalized performance indicators for comparative purposes, rather than as quantities directly derived from the absolute radiated power and thermal noise levels provided by full-wave electromagnetic simulations. 
In this perspective, the infinite and finite ground-plane configurations should be interpreted as two limiting cases, corresponding, respectively, to ideal total isolation and to a weakly isolated, fully open SIM structure.

Once the S-matrix is computed, the optimal set of parameters
$\boldsymbol{\eta}$ is obtained using the optimization method described
in Section~\ref{sec:SIM_model}. The resulting optimal parameter set
defines the corresponding matrix $\bm{\Gamma}(\boldsymbol{\eta})$,
which can be directly implemented in the electromagnetic simulator as a
SPICE two-port network connecting all the unit cells of the SIM layers.

Two different types of SPICE two-port networks are considered in the
examples: ideal lossless networks implementing continuous phase
shifting, and non-ideal lossy networks implementing quantized phase
shifting. In the following, the parameter $P$ denotes the number of admissible discrete
states of the phase shifter.

In the first case, ideal two-port networks allow arbitrary phase shifts
since the phase is treated as a continuous optimization variable. This
limiting case is denoted by $P=\infty$, corresponding to continuous phase 
control, and provides an upper bound on
the achievable SIM performance. 
It allows isolating the impact of the phase configuration from hardware impairments.

In the second case, to obtain a more realistic performance assessment,
the two-port networks are derived from measured phase responses of a
practical off-the-shelf phase shifter. Specifically, we consider the
Qorvo TGP2100 5-bit digital MMIC phase shifter\footnote{\url{https://www.qorvo.com/products/p/TGP2100}}.  The availability of
measured S-parameters allows simulations with
$P \in \{32,16,8,4\}$ discrete states. The corresponding two-port
networks exhibit non-ideal effects including insertion loss
(typically $\cong 18$~dB) and finite return loss
(typically $\cong 6$~dB), which are explicitly accounted for in the
full-wave co-simulation.

Finally, a numerical full-wave co-simulation is performed to obtain the
end-to-end S-matrix of the considered scenario.
\color{black}
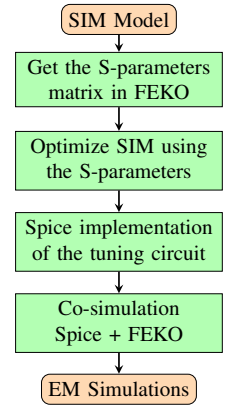
\begin{figure}
    \centering
    \scalebox{0.8}{
\begin{tikzpicture}[node distance=2cm]
\node (start) [startstop] {SIM Model};
\node (pro1) [process, below of=start, yshift= 1 cm] {Get the S-parameters matrix in FEKO};
\node (pro2) [process, below of=pro1, yshift= 0.65 cm] {Optimize SIM using the S-parameters};
\node (pro3) [process, below of=pro2, yshift= 0.65 cm] {Spice implementation of the tuning circuit };
\node (pro4) [process, below of=pro3, yshift= 0.65 cm] {Co-simulation Spice + FEKO};
\node (stop) [startstop, below of=pro4, yshift= 0.9 cm] {EM Simulations};
\draw [arrow] (start) -- (pro1);
\draw [arrow] (pro1) -- (pro2);
\draw [arrow] (pro2) -- (pro3);
\draw [arrow] (pro3) -- (pro4);
\draw [arrow] (pro4) -- (stop);
\end{tikzpicture}}
    \caption{Algorithmic pipeline used to obtain and validate the end-to-end results.}
    \vspace{-2mm}
    \label{fig:pipeline}
\end{figure}
All steps of the SIM optimization described above are visually summarized in Fig. \ref{fig:pipeline}.

\subsection{Communication-oriented scenarios}
\label{sec:comm_scenarios}

In this section, we investigate communication-oriented scenarios aimed at exploiting the reconfigurability of a SIM to orthogonalize multi-stream wireless channels. Two related scenarios are considered, namely \emph{COMM-MIMO} and \emph{COMM-MU-SIMO}. Although motivated by different
application settings, both scenarios can be cast into a unified mathematical
problem, which allows a common analysis and performance evaluation.
As mentioned, in order to keep the computational complexity of the electromagnetic simulations within manageable limits, the SIM is implemented with a limited number of elements. 
Specifically, we considered a SIM configuration with $N_y^q=16$, $N_z^q=4$, and $Q=5$ layers, 
corresponding to a total of  $Q N_y^q N_z^q = 320$ tunable parameters.
Both the case with infinite and finite ground plane have been considered. In both this two cases simulations are carried out with $P\in [\infty,32,16,8,4]$. 

\begin{figure}[h]
    \centering
    \includegraphics[width=1\columnwidth]{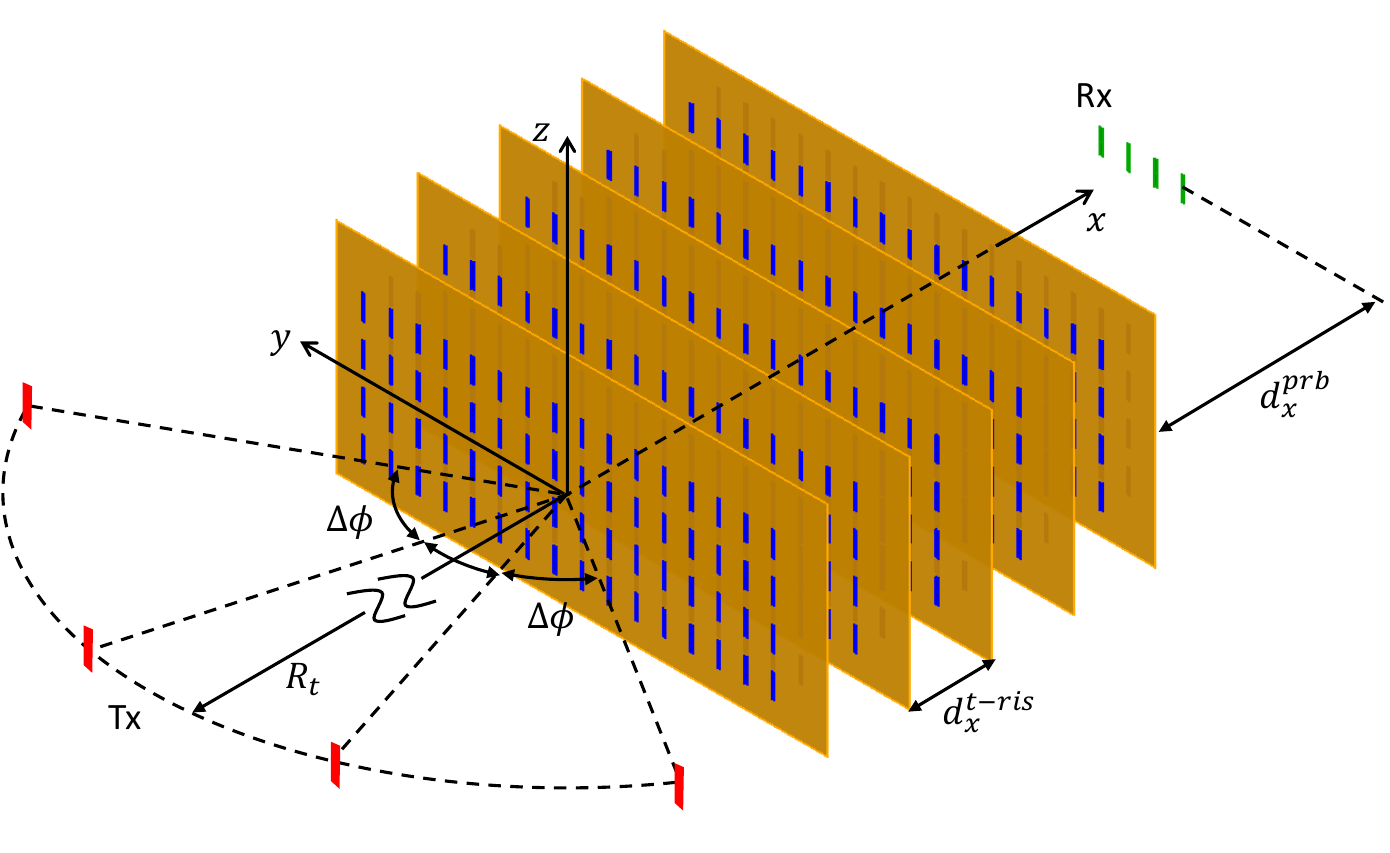}
    \caption{COMM-MU-SIMO scenario.}
    \vspace{-2mm}
    \label{fig:comm_mu_simo_scn}
\end{figure}

\paragraph{COMM-MU-SIMO}
The first scenario is inspired by the multi-user MISO setup studied
in~\cite{10922857}, where multiple user terminals are located at given distances from the SIM and from each other, and the SIM is configured to orthogonalize the channels between a transmitting array placed in proximity of the SIM and the
receivers. 
To maintain notational and conceptual uniformity, we consider here the reciprocal \emph{MU-SIMO} configuration in which the receiver is an array located close to the SIM, while the transmitters correspond to spatially separated nodes far from the SIM. A schematic graphical representation
of the this scenarios is reported in Fig. \ref{fig:comm_mu_simo_scn}.

Specifically, in this configuration, four transmitters are placed on a circle 
at a distance $R_t = 220\lambda$ from the origin, which is located at the center 
of the first ground plane. Therefore, the system operates under strong far-field 
conditions. The transmitters are uniformly spaced in angle with a separation 
$\Delta\phi = 12^{\circ}$, and are separated by several wavelengths. This 
configuration represents four independent and clearly decoupled transmitting 
nodes, as in the scenario reported in~\cite{10922857}.

The receiver probes are arranged in a UPA configuration with 
$N_y^{prb} = 4$ and $N_z^{prb} = 1$, as described in 
Section~\ref{subsec:simulation_setup}. The array is perfectly centered 
with respect to the last layer of the SIM and is located at a distance 
$d_x^{prb} = 4\lambda$ from it.

\begin{figure}[h]
    \centering
    \includegraphics[width=1\columnwidth]{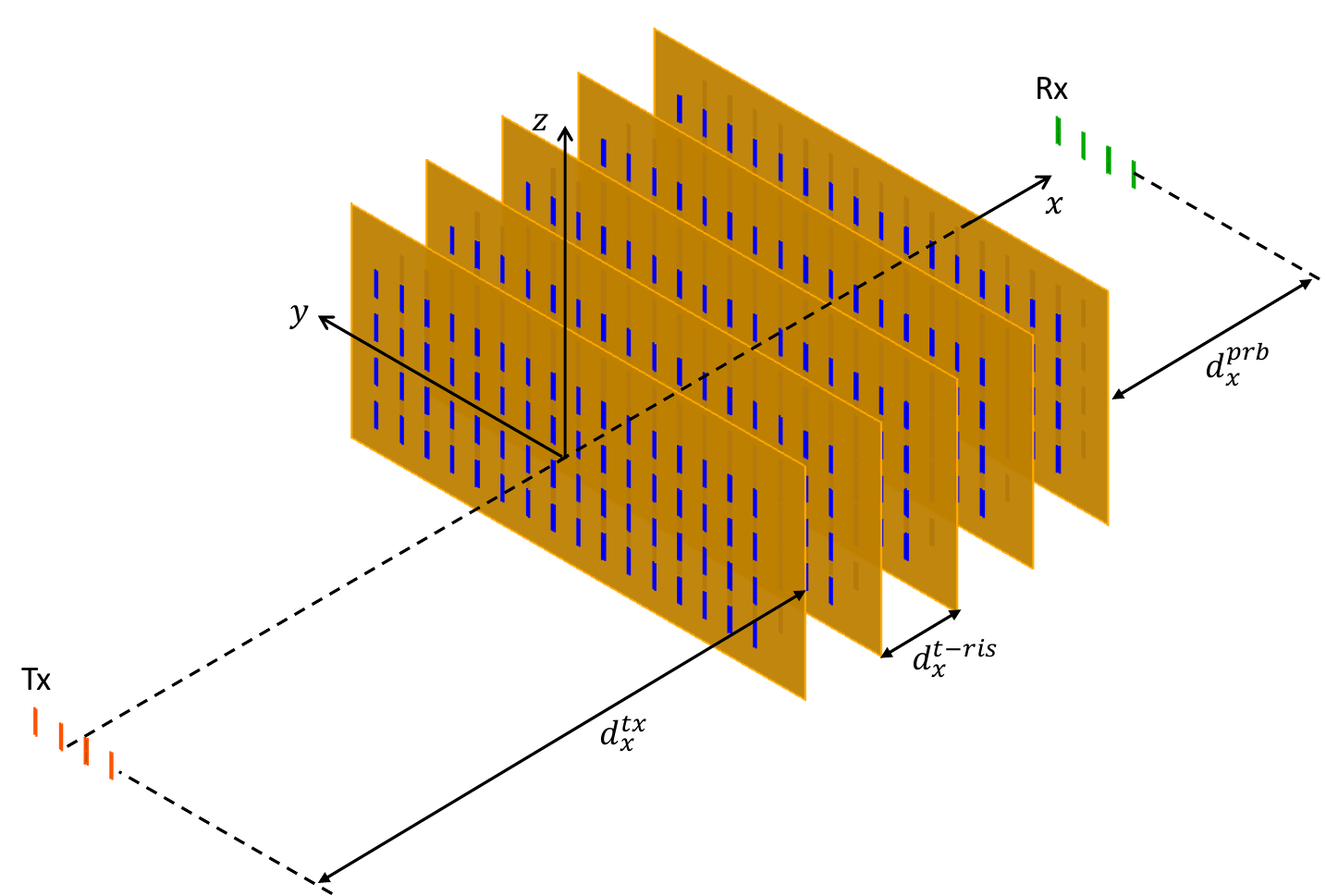}
    \caption{COMM-MIMO scenario.}
    \vspace{-2mm}
    \label{fig:comm_mimo_scn}
\end{figure}

\paragraph{COMM-MIMO}
The second scenario, referred to as COMM-MIMO, is inspired by the setup
considered in~\cite{10158690}, where two SIMs are employed, one at the
transmitter side and one at the receiver side, to orthogonalize a MIMO channel
and maximize its capacity. In the present work, we consider a simplified yet
fundamental variant of this problem, in which a \emph{single} SIM is used to
orthogonalize the end-to-end channel. In this case, channel diagonalization can
be achieved through a zero-forcing–type configuration, provided that the
SIM-assisted channel admits full rank. 
A schematic graphical representation
of the this scenarios is reported in Fig. \ref{fig:comm_mimo_scn}.

In the considered setting, the four receiver are arranged as in the  previous scenario and the transmitters are arranged according to the
same UPA geometry of the receiver, but placed at a distance of $d_x^{tx}=10\lambda$ from the SIM, thus operating under  near-field
conditions, where the resulting channel generally exhibits full rank (i.e.,
rank~$4$). In this case, the scenario is inherently more challenging than the
previous one, as it requires the orthogonalization of a MIMO channel, similarly
to the problem considered in~\cite{10158690}.


\subsection{Unified formulation and feasibility conditions}

Consider $L$ transmit ports and $L$ receive ports, and collect $I=L$ canonical
input excitations as
\[
\bm A_S = \bm I_L .
\]
The corresponding output matrix is given by
\begin{equation}
\bm Y
=
\bm S_{RT}
\;+\;
\bm G\,\bm T(\boldsymbol{\eta})\,\bm F ,
\label{eq:Y_with_direct}
\end{equation}
where $\bm F$ and $\bm G$ denote the geometry-induced operators associated with
the TX$\rightarrow$SIM and SIM$\rightarrow$RX links, respectively.

The communication objective is to achieve channel diagonalization, i.e.,
\begin{equation}
\bm Y \;\approx\; \alpha\,\bm I_L,
\qquad \alpha\in\mathbb C,
\label{eq:identity_target_with_direct}
\end{equation}
which corresponds to suppressing inter-stream interference up to a complex scaling factor.

\subsection{Performance metrics}

For both COMM-MIMO and COMM-MU-SIMO scenarios, performance is assessed using two
complementary metrics. First, we provide a graphical representation of the
degree of channel orthogonalization achieved by the optimized SIM. Second, we
evaluate the average per-channel capacity, computed by modeling the residual
inter-stream interference as additive white noise and considering different
signal-to-noise ratio (SNR) levels. The SNR is defined as the ratio between the
per-stream signal power at the SIM output and the noise power.

To quantitatively assess the communication performance, we consider the
aggregate spectral efficiency achieved after SIM optimization.
Specifically, for a given configuration, the performance metric is defined as
\begin{equation}
C
=
\sum_{\ell=1}^{L}
\log_2\!\left(1+\mathrm{SINR}_\ell\right),
\label{eq:C_def}
\end{equation}
where $\mathrm{SINR}_\ell$ denotes the signal-to-interference-plus-noise ratio
of the $\ell$-th stream, computed by considering as useful signal the
aggregate received signal power (i.e., the sum of the desired signal
contributions). For each stream, the SINR is computed by modeling the residual inter-stream
interference due to imperfect channel diagonalization as additive noise.
Accordingly, $\mathrm{SINR}_\ell$ accounts for both the thermal noise component
and the interference leakage from the remaining streams, and is expressed as
the ratio between the useful signal power at the SIM output for stream $\ell$
and the sum of the noise power and the interference power affecting that
stream. The noise power is normalized with respect to the average received
signal power per stream at the SIM output.

Using this definition, we consider the following reference and SIM-assisted
performance measures:
\begin{itemize}
\item $C_{\mathrm{SIM-I}}^{(P)}$, obtained with an SIM optimized according
to the Isolated SIM model;
\item $C_{\mathrm{SIM-W}}^{(P)}$, obtained with an SIM optimized according
to the Weakly-isolated SIM model;
\item $C_{\mathrm{SIM-NI}}^{(P)}$, obtained with an SIM optimized using the
fully general multiport model.
\end{itemize}

In all cases, the parameter $P$ denotes the number of admissible discrete states of each unit cell. 
We now present the results related to the communication-oriented (COMM) scenarios. 
We start from the \emph{COMM-MU-SIMO} case by considering infinite ground planes, and we report the corresponding results obtained by optimizing
the SIM according to the \emph{SIM-I} model.

\begin{figure}[t]
\centering
\includegraphics[width=1\linewidth]{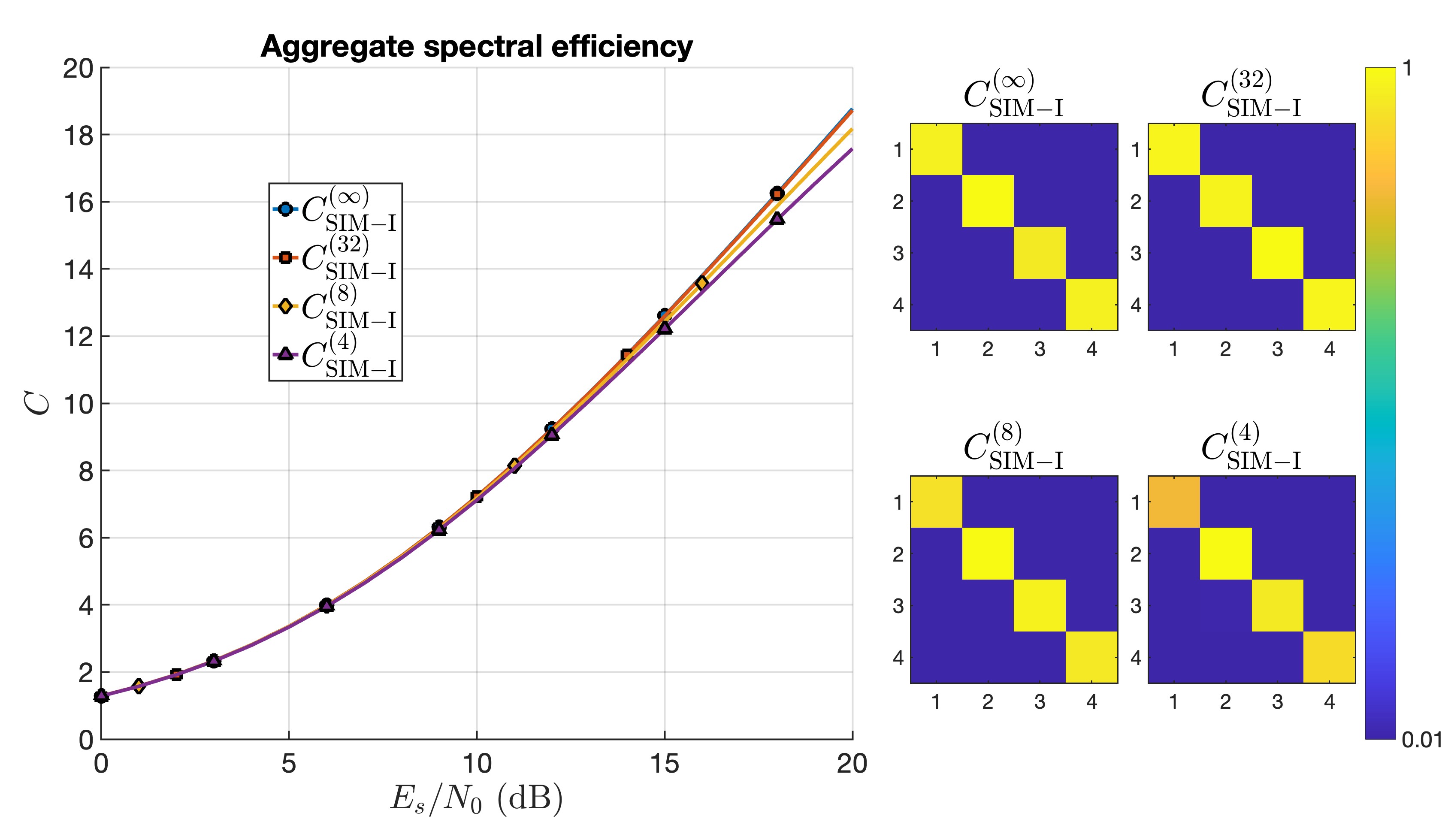}
\caption{COMM-MU-SIMO scenario with infinite ground planes: channel
orthogonalization performance obtained with the SIM-I model.}
\vspace{-2mm}
\label{fig:comm_mu_simo_gnd_inf}
\end{figure}

\begin{figure}[t]
\centering
\includegraphics[width=1\linewidth]{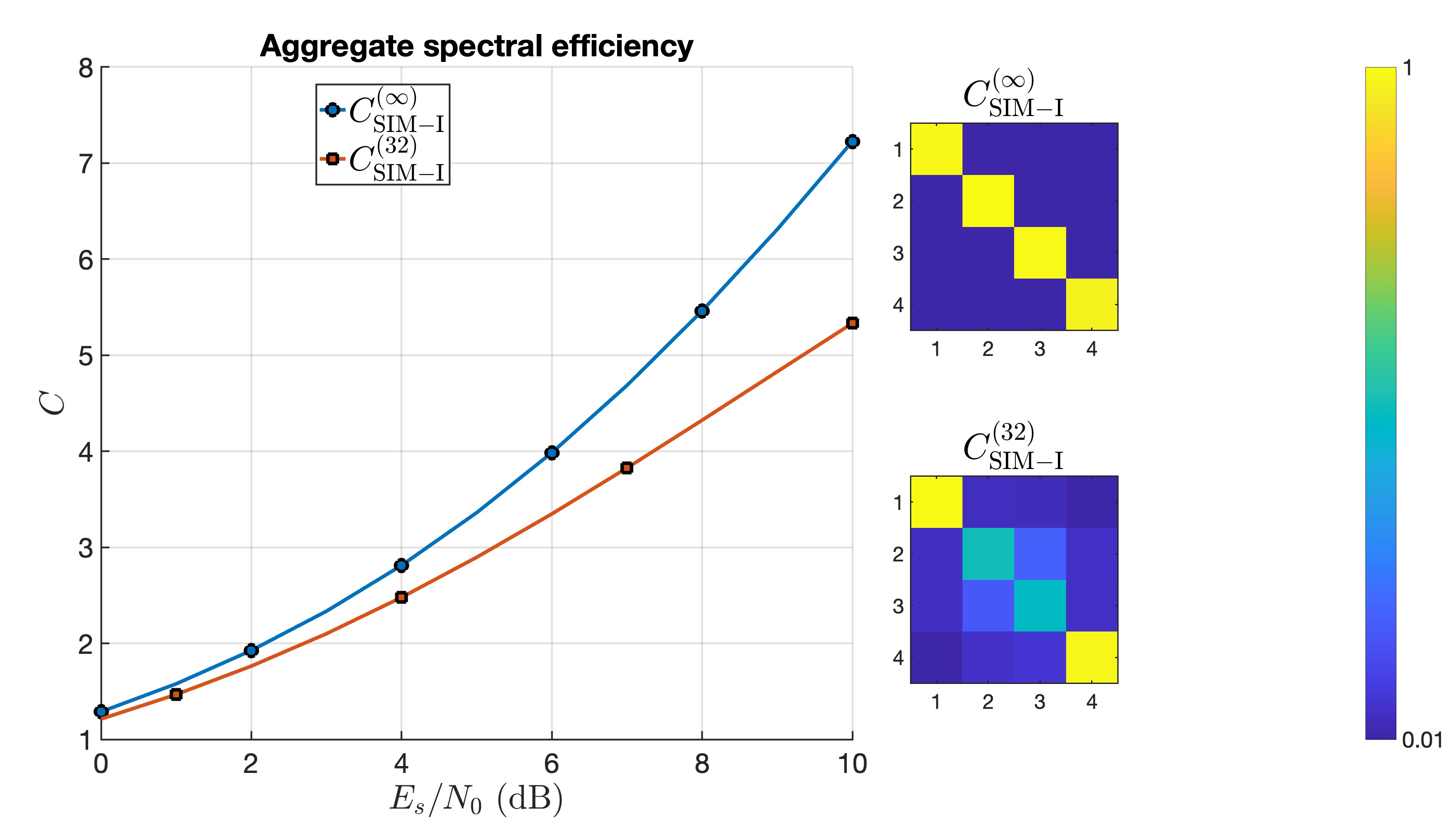}
\caption{COMM-MIMO scenario with infinite ground planes: channel
orthogonalization performance obtained with the SIM-I model.}
\vspace{-2mm}
\label{fig:comm_mimo_gnd_inf}
\end{figure}

\begin{figure}[t]
\centering
\includegraphics[width=\linewidth]{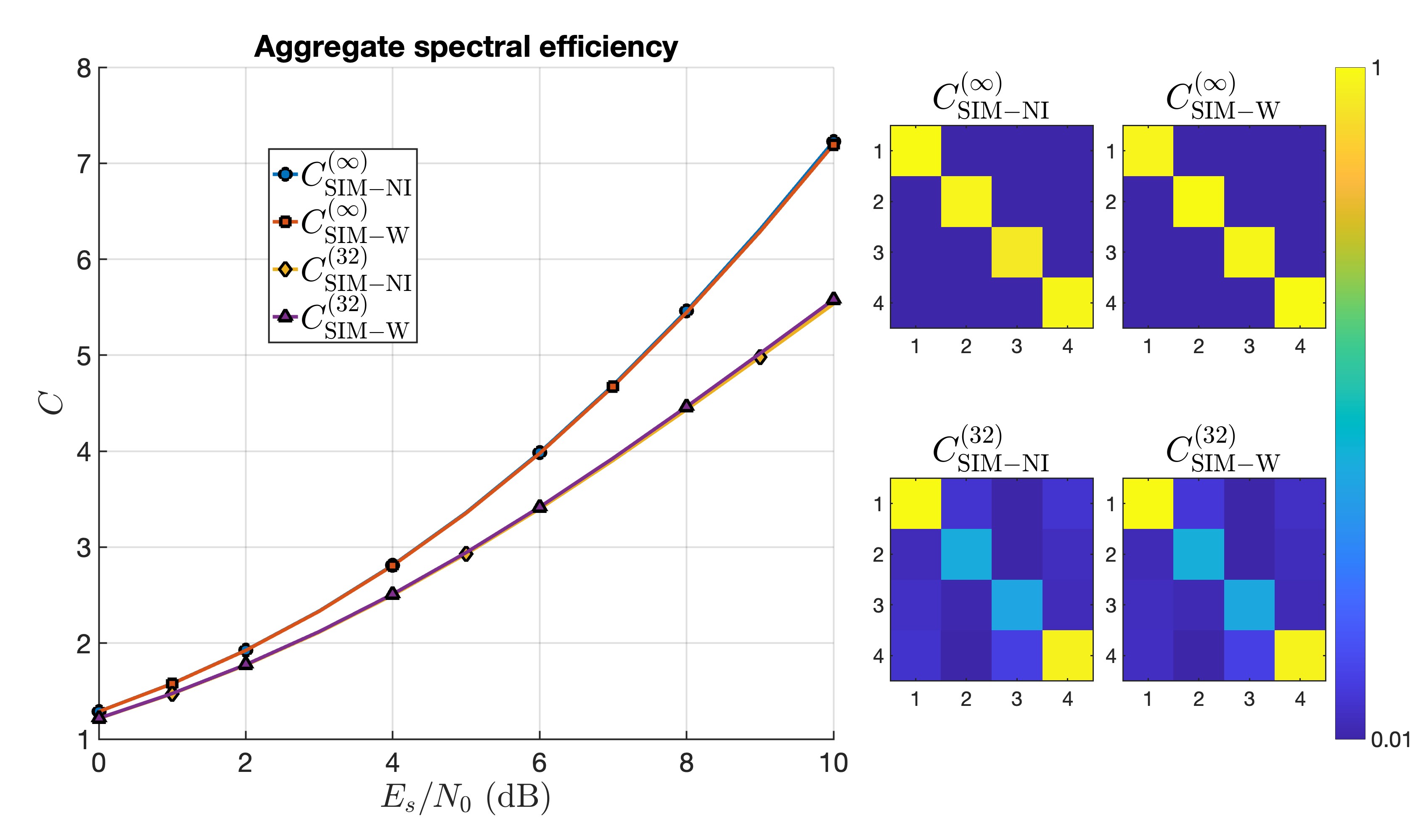}
\caption{COMM-MU-SIMO scenario with finite ground planes: channel
orthogonalization performance obtained with the SIM-I model.}
\vspace{-2mm}
\label{fig:comm_mu_simo_gnd_fin}
\end{figure}

\begin{figure}[t]
\centering
\includegraphics[width=1\linewidth]{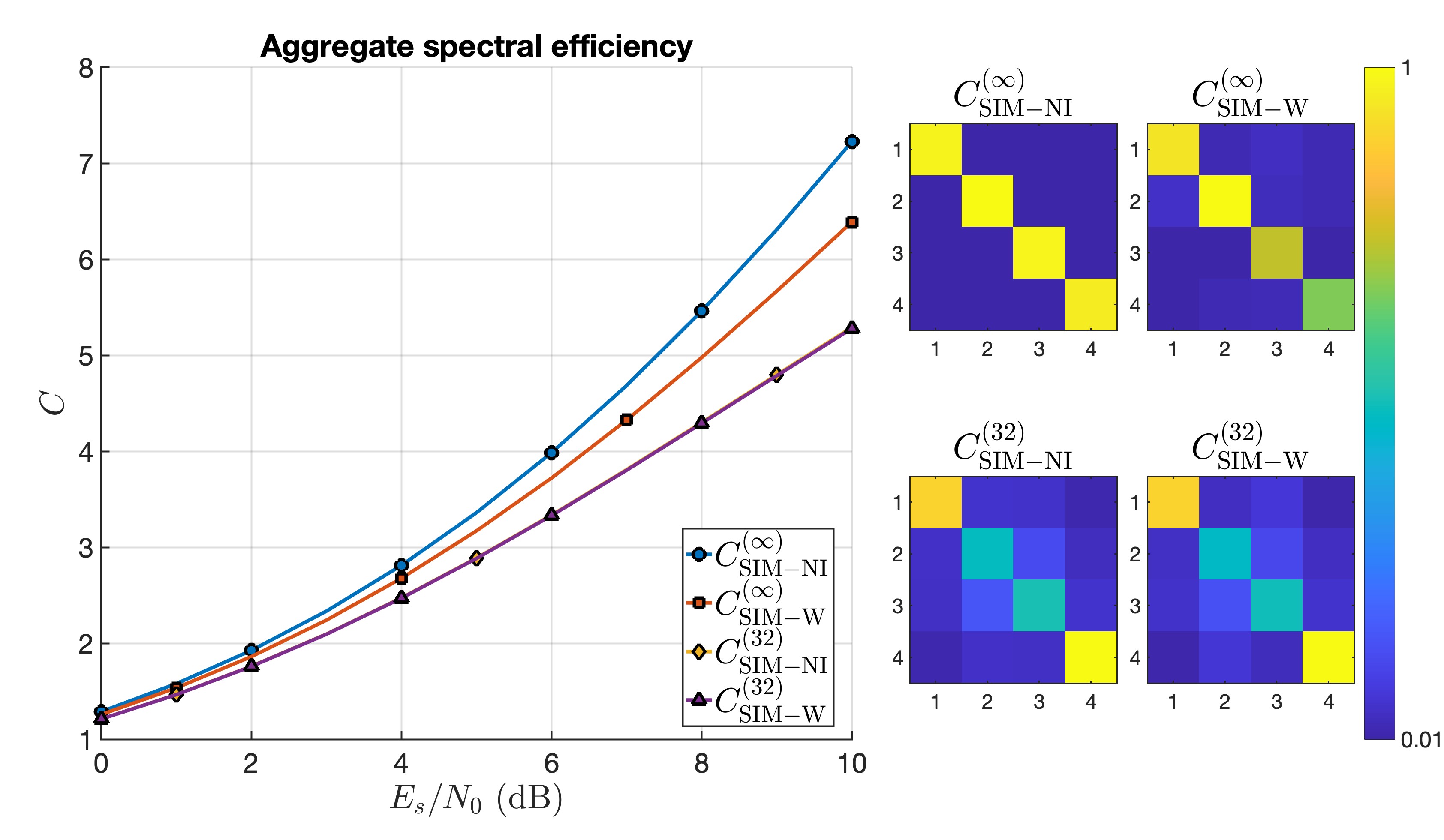}
\caption{COMM-MIMO scenario with finite ground plane: comparison between the SIM-W and SIM-NI optimization approaches.}
\vspace{-2mm}
\label{fig:comm_mimo_gnd_fin}
\end{figure}

In Fig.~\ref{fig:comm_mu_simo_gnd_inf}, we illustrates the degree of channel orthogonalization achieved in the COMM-MU-SIMO setup. For $P=\infty$ (continuous phase control), the SIM-I architecture achieves near-perfect orthogonalization, with the interference terms effectively suppressed. 
When practical phase shifts with finite resolution are considered, the
performance remains very good even with a small number of quantization levels
(e.g., $P=4$). A noticeable capacity loss appears only at high SNR, where the
residual interference is no longer masked by thermal noise and becomes the
dominant impairment. Notably, this is the only scenario in which satisfactory
performance can be achieved even with coarse phase quantization. For this
reason, in the following figures we report only the case $P=32$.

Figure~\ref{fig:comm_mimo_gnd_inf} quantifies the degree of channel
orthogonalization achieved in the COMM-MIMO setup. Under continuous phase
control ($P=\infty$), the SIM-I architecture attains a near-ideal
diagonalization of the MIMO channel, with the interference components almost completely suppressed 
When phase quantization is introduced, even with
$P=32$ levels, residual off-diagonal leakage appears and becomes increasingly
relevant at high input SNR. This behavior stems from the fact that MIMO channel
orthogonalization is inherently more sensitive to phase perturbations:
achieving the desired zero-forcing effect requires phase shifts with very
high precision and resolution.

In Fig. \ref{fig:comm_mu_simo_gnd_fin} and Fig. \ref{fig:comm_mimo_gnd_fin}, we
compare the optimization outcomes obtained with the SIM-NI and SIM-W models
when finite ground planes introduce non-negligible inter-layer (inter-block) coupling. In the
COMM-MU-SIMO scenario, both models achieve good channel orthogonality under
continuous-phase control. Indeed, the SIM-W
model attains almost the same performance as SIM-NI, indicating
that the weak-coupling approximation is effective for MU-SIMO orthogonalization.

Conversely, in the COMM-MIMO scenario, where achieving channel diagonalization in the near-field is more challenging, the SIM-W model exhibits a performance degradation with respect to SIM-NI. Finally, when phase quantization is considered, the results obtained with SIM-W and SIM-NI are very
similar, and both exhibit a significant degradation compared to the
continuous-phase case.

\FloatBarrier

\subsection{Sensing-oriented scenarios}
\label{sec:sensing_scenarios}

In this section, we consider sensing-oriented scenarios in which the goal
is to perform a joint \emph{range--angle} estimation in a 2D geometry from the
signal received from an unknown transmitter. Our approach builds upon the
framework proposed in~\cite{abrardo_Eusipco}, where the SIM is exploited to
induce distinctive output signatures associated with different transmitter
locations.

Following~\cite{abrardo_Eusipco}, we define a polar grid of candidate
transmitter locations with variable range and angle. Specifically, the grid is
constructed according to the criterion given in~\cite{abrardo_Eusipco} so as to balance the
range and angular resolution while keeping the number of test points limited.
For each grid point, the SIM outputs are analyzed to determine the most likely
location of the transmitter, i.e., the estimate is obtained by selecting the
grid point whose output signature best matches the observed one. For positions
lying between grid points, refined estimates can be obtained through suitable
interpolation functions, as discussed in~\cite{abrardo_Eusipco}.

The SIM is optimized by enforcing a desired output response for each grid
location. Hence, the design methodology is consistent with the COMM scenarios
considered earlier and is formulated as the minimization of a loss function
measuring the mismatch between the SIM output matrix and a prescribed target.

Due to the computational burden of electromagnetic simulations in FEKO, we split the joint range--angle estimation task into two separate subproblems, for the aim to keep the electromagnetic simulation effort manageable while preserving the ability
to assess the sensing performance of the optimized SIM. 

\begin{figure}[h]
    \centering
    \includegraphics[width=1\columnwidth]{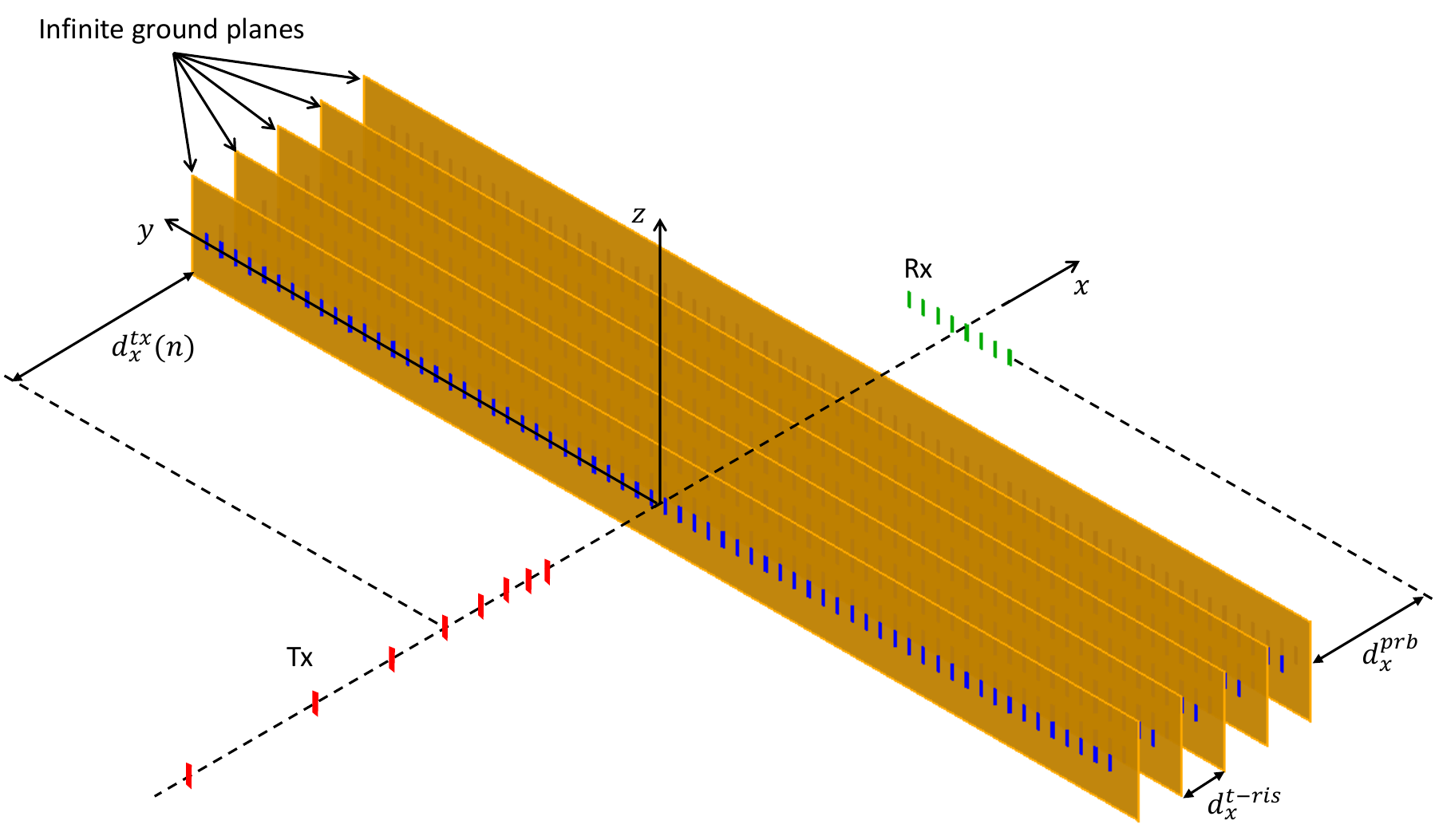}
    \caption{Range estimation scenario.}
    \vspace{-2mm}
    \label{fig:range_estim}
\end{figure}

\begin{figure}[h]
    \centering
    \includegraphics[width=1\columnwidth]{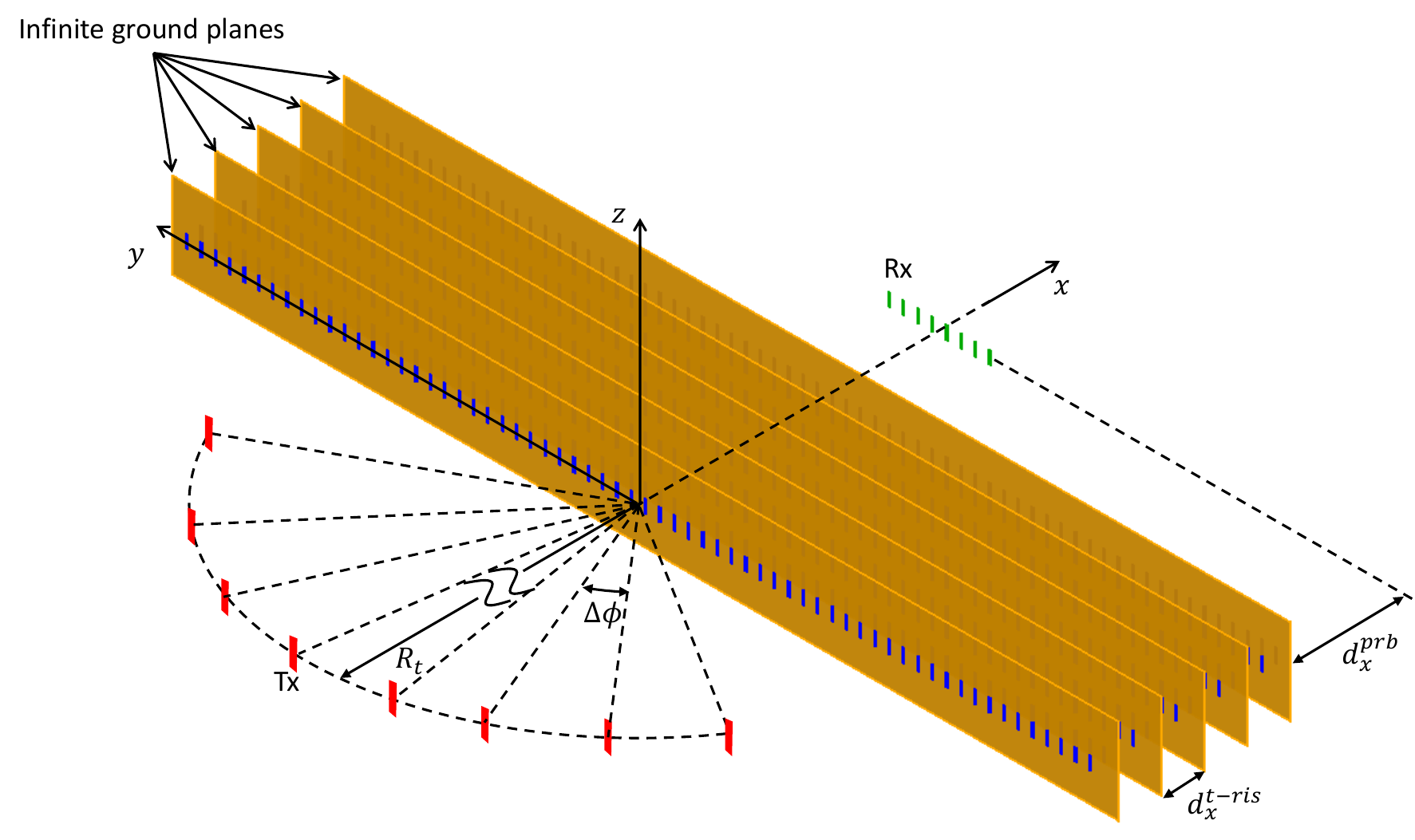}
    \caption{DOA estimation scenario.}
    \vspace{-2mm}
    \label{fig:doa_estim}
\end{figure}

{In Fig. \ref{fig:range_estim} the geometry configuration of the first subproblem is shown. In this case, distance estimation is considered for a fixed angle 
$\theta = 0$. As shown in~\cite{abrardo_Eusipco}, this leads to a radial grid defined by points of the form $R_{\max}/n$, i.e., located at 
non-uniform distances.
Figure~\ref{fig:range_estim} shows the considered configuration for 
$N_t = 8$ grid points, which correspond to the same number of transmitters in the FEKO simulations.
The receiver probes are arranged in the same UPA configuration of the first subproblem.} In Fig.~\ref{fig:doa_estim}, the geometry configuration of the second subproblem is illustrated. 
The objective in this case is angular (DoA) estimation. Given the presence of an SIM with 64 elements, 
the optimal angular sampling is defined as
\begin{equation}
\theta_n = \arcsin\!\left(\frac{n}{32}\right).
\end{equation}
Since $N_t = 8$ grid points are selected for the transmitters, the resulting angular span is 
approximately $15^\circ$. The transmitters are placed on a circle in the $xy$-plane at a distance 
$R_t = 1000\lambda$ from the origin, which is located at the center 
of the first ground plane, with an almost uniform angular separation of 
approximately $\Delta\phi = 2^\circ$.
The $N_r=8$ receiver probes are arranged in a UPA configuration with 
$N_y^{prb} = 8$ and $N_z^{prb} = 1$, as described in 
Section~\ref{subsec:simulation_setup}. The array is perfectly centered with respect to the last layer of the SIM and is located at a distance $R_t$.


In both subproblems, the SIM is configured with $Q=5$ layers, and the strip dipoles are arranged in a ULA configuration with $N_y^q=64$, $N_z^q=1$ 
with $d_y^q=\lambda/2$, ($1D$ aperture). This choice aims at maximizing the physical aperture while
avoiding an increased implementation complexity in FEKO associated with a large
number of elements along both dimensions.  The resulting maximum dimension of aperture is
\[
D \approx (N_y^q/2)\lambda = 32\lambda .
\]
At $f_0=28$~GHz (i.e., $\lambda=0.0107$~m), this yields $D\approx 0.342$~m
(about $34$~cm). The corresponding far-field (Fraunhofer) distance is
\[
d_{\mathrm{FF}} \approx \frac{2D^2}{\lambda} \approx 22~\text{m}.
\]
Since accurate range estimation requires near-field conditions (i.e., a
pronounced wavefront curvature across the aperture), we focus on transmitter
ranges well below $d_{\mathrm{FF}}$, and in particular on distances smaller
than $1$~m. Conversely, the angular estimation does not impose the same
constraint and can be meaningfully investigated over a wide angular span.

\paragraph{Performance metrics and noisy observations}
To assess sensing performance, we consider the estimation error of the target
parameter, namely the \emph{range} error or the \emph{angle} error, depending on
the specific subproblem under investigation. Although the SIM is trained on a
finite polar grid of candidate locations, performance is evaluated over a much
larger set of test points spanning the entire area of interest.

To enable estimation at arbitrary positions (i.e., not restricted to the grid),
we assume that the SIM outputs are post-processed through a polynomial
interpolator. Specifically, the interpolator fits the SIM output signatures
obtained at the grid points and provides a continuous mapping that can be
queried to estimate the transmitter position (or equivalently the corresponding
range/angle) for any point within the region of interest. The estimation error
is then computed over a dense set of representative test locations.

Furthermore, we assume that the received SIM outputs are affected by additive
white noise. Accordingly, the observation model used for performance evaluation
is
\begin{equation}
\bm y = \bm y_{\mathrm{clean}} + \bm n,
\qquad
\bm n \sim \mathcal{CN}(\bm 0, \sigma_n^2 \bm I),
\label{eq:sensing_noise_model}
\end{equation}
where $\bm y_{\mathrm{clean}}$ denotes the noiseless SIM output vector
associated with the considered transmitter location. The noise variance
$\sigma_n^2$ is set to match a prescribed SNR, defined with respect to the
average useful received power across the considered output ports. 

The estimation accuracy is quantified in terms of the standard deviation of the
estimation error of the parameter of interest, namely the range error
$\sigma_d$ or the angle error $\sigma_\theta$, depending on the considered
scenario. Specifically, given a set of test locations spanning the region of
interest, the estimation error is computed for each test point and the
corresponding standard deviation is evaluated over the entire set.

We consider SIM configurations with infinite ground planes,
due to the significantly higher computational burden associated with finite
ground-plane models. Accordingly, all results are obtained using the SIM-I
model and are reported for different levels of phase discretization $P$. Specifically, we evaluate the standard deviation of the estimation error for
the \emph{range} and \emph{angle} estimation subproblems separately, denoted as
$\sigma_d^{(P)}$ and $\sigma_\theta^{(P)}$, respectively.

\begin{figure}[t]
\centering
\includegraphics[width=\linewidth]{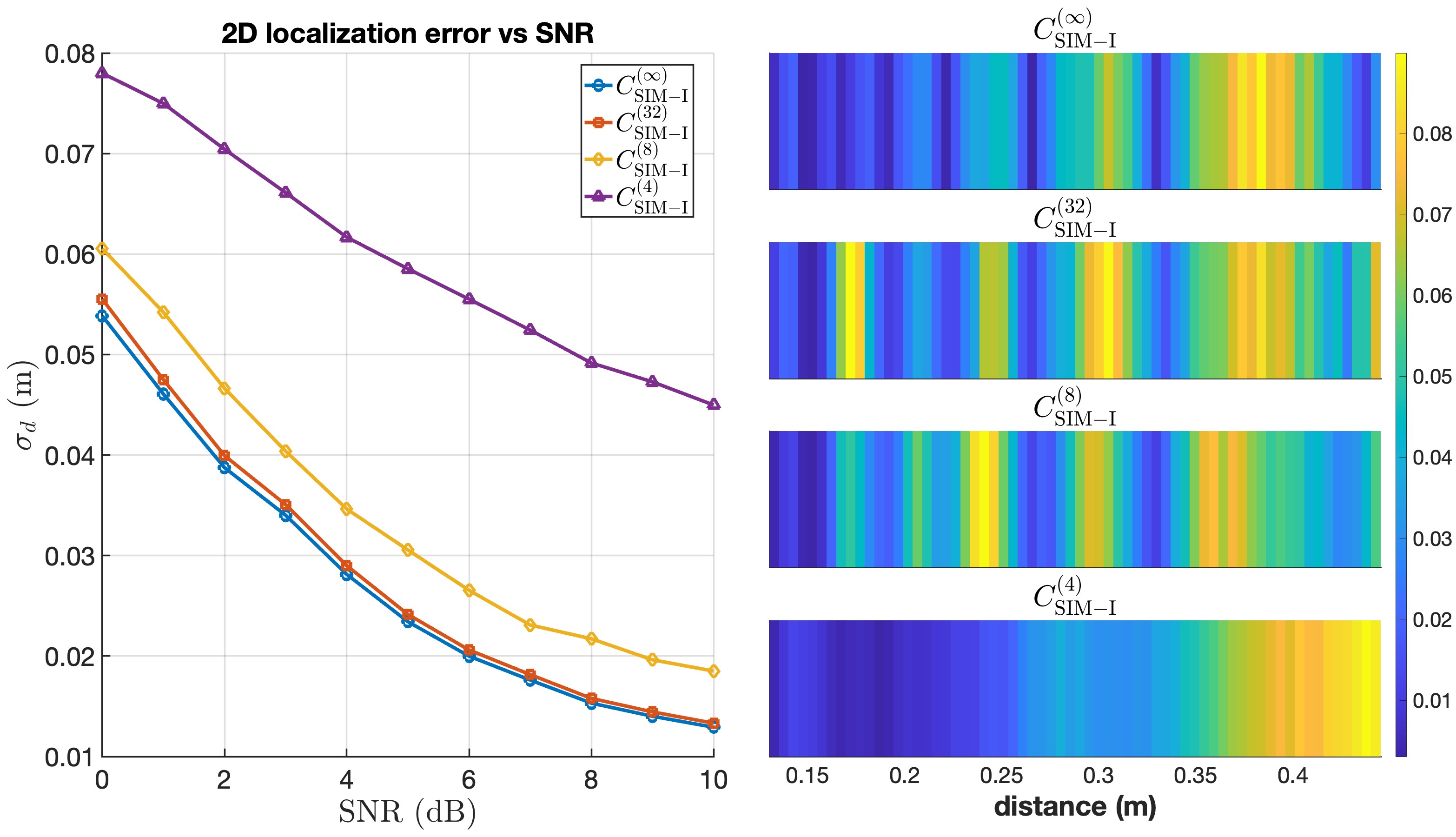}
\caption{Range-estimation performance for the SIM-I model: standard deviation $\sigma_d^{(P)}$ of the distance-estimation error for different phase quantization levels $P$, as a function of the SNR evaluated at the farthest test points). The false-color bars report the empirical distribution of the estimation error as a function of the transmitter distance
for the corresponding cases.}
\vspace{-2mm}
\label{fig:loc_distance_simid}
\end{figure}

Figure~\ref{fig:loc_distance_simid} shows the distance-estimation accuracy
($\sigma_d$) obtained by optimizing the SIM over a grid of transmitter ranges
and applying polynomial interpolation at the test points. As expected,
continuous-phase control ($P=\infty$) achieves the lowest values of
$\sigma_d$ over the entire SNR range. However, it can be observed that the use
of quantized phase shifts with $P$ discrete levels still enables acceptable
performance, provided that the number of quantization levels is not smaller
than $P=8$.

The false-color representation highlights, for the case $P=\infty$, eight
pronounced dark horizontal bands corresponding to the grid points used during
the SIM training phase. These grid locations naturally yield the smallest
distance-estimation errors.

\begin{figure}[t]
\centering
\includegraphics[width=\linewidth]{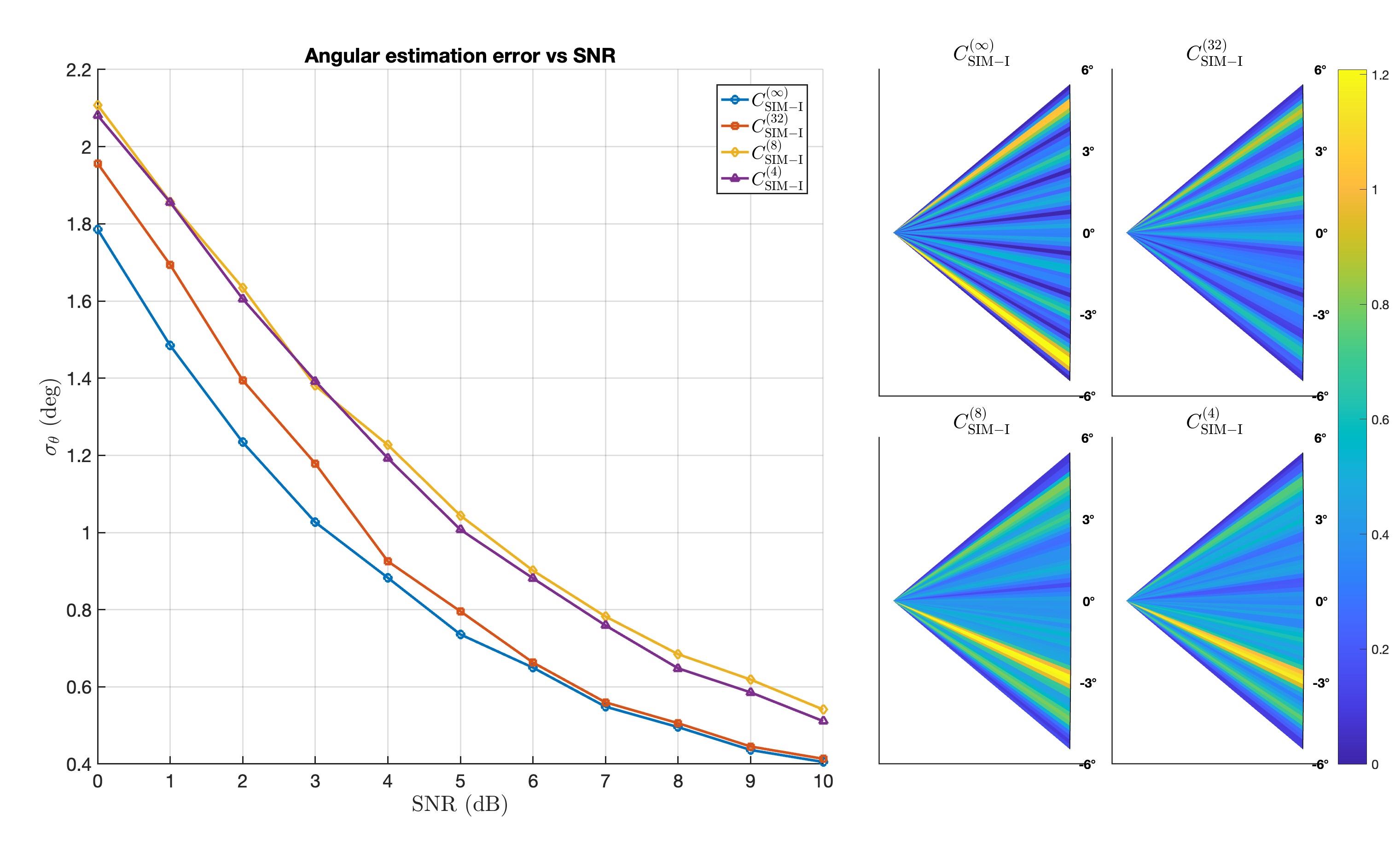}
\caption{Angle-estimation performance for the SIM-I model: standard deviation
$\sigma_\theta^{(P)}$ of the angle-estimation error for different phase
quantization levels $P$, as a function of the SNR (defined as the SNR measured
at the farthest test points). The false-color bars report the empirical
distribution of the estimation error as a function of the transmitter angle,
represented in a cone-like view centered at the SIM position, for the
corresponding cases.}
\vspace{-2mm}
\label{fig:loc_angle_simid}
\end{figure}

Figure~\ref{fig:loc_angle_simid} reports the angular-estimation accuracy
($\sigma_\theta$) for the SIM-I model, using the same phase-quantization levels
and SNR sweep adopted in Fig.~\ref{fig:loc_distance_simid}. In this case, it can
be noted that angle estimation is inherently less challenging than range
estimation; consequently, good performance is achieved even with coarse phase
quantization, e.g., $P=4$. As before, the false-color representation clearly
highlights the grid points used for SIM training, which in this case correspond
to eight distinct angular directions.

\section{Conclusion}

This paper presented a multiport network-theoretic framework for modeling and optimizing stacked intelligent metasurfaces while preserving electromagnetic consistency. By adopting an S-parameter representation, the proposed formulation captures multiport interactions, tunable circuit effects, and hardware non-idealities within a unified optimization framework. Three SIM models with different levels of electromagnetic coupling were considered, highlighting the trade-off between modeling accuracy and computational complexity. The optimized configurations were validated through electromagnetic simulations, confirming the reliability of the proposed approach. Results obtained in both communication and sensing scenarios show that SIMs can effectively manipulate the electromagnetic channel, achieving channel orthogonalization and enabling localization functionalities even with practical phase quantization. Overall, the proposed methodology provides a useful bridge between optimization-oriented models and realistic electromagnetic implementations of SIM-based systems.

\section*{Acknowledgment}
The work of Andrea Abrardo and Alberto Toccafondi was supported by the EU -- NextGenerationEU under the National Recovery and Resilience Plan (PNRR), through the Cascade Funding Call of Spoke 7 ``GREEN AND SMART ENVIRONMENTS,'' within the project ``Smart Metasurfaces Advancing Radio Technology (SMART).''

This work was conducted when G. Pettanice was with Universite Paris-Saclay, CNRS, CentraleSupelec, Laboratoire des Signaux et Systemes. During this time, the work of G. Pettanice was supported by the Agence Nationale de la Recherche (ANR) through the France 2030 project ANR-PEPR Networks of the Future under grant agreement NF-SYSTERA 22-PEFT-0006, and in part by the European Union through the Horizon Europe project COVER under grant agreement number 101086228.

The work of M. Di Renzo was supported in part by the European Union through the Horizon Europe project COVER under grant agreement number 101086228, the Horizon Europe project UNITE under grant agreement number 101129618, the Horizon Europe project INSTINCT under grant agreement number 101139161, and the Horizon Europe project TWIN6G under grant agreement number 101182794, as well as by the Agence Nationale de la Recherche (ANR) through the France 2030 project ANR-PEPR Networks of the Future under grant agreement NF-SYSTERA 22-PEFT-0006, and by the CHIST-ERA project PASSIONATE under grant agreements CHIST-ERA-22-WAI-04 and ANR-23-CHR4-0003-01. Also, the work of M. Di Renzo was supported in part by the Engineering and Physical Sciences Research Council (EPSRC), part of UK Research and Innovation, and the UK Department of Science, Innovation and Technology through the CHEDDAR Telecom Hub under grant EP/X040518/1 and grant EP/Y037421/1, and through the HASC Telecom Hub under grant EP/X040569/1.

\appendices
\section{Computation of the Full Inverse Matrix}
\label{app:full_T}

In this appendix we describe the block-recursive procedure used to compute
the full inverse matrix
\begin{equation}
\mathbf T
\;\triangleq\;
\left(\bm\Gamma^{-1}-\bm S_{EE}^{(0)}\right)^{-1}
\in\mathbb C^{(2QK)\times(2QK)},
\label{eq:T_def_appendix}
\end{equation}
which arises in both the $SIM-I$ and $SIM-NI$ formulations.
The derivation follows the same principles introduced in
\cite{Abra_SIM1} and exploits the banded block structure induced by the
layered SIM architecture.
Let
\[
\mathbf S
\;\triangleq\;
\bm\Gamma^{-1}-\bm S_{EE}^{(0)} ,
\]
and partition $\mathbf S$ into $2Q\times 2Q$ blocks of size $K\times K$ as
\begin{equation}
\mathbf S
=
\begin{bmatrix}
\mathbf B_1 & \mathbf C_1 & \mathbf 0 & \cdots & \mathbf 0\\
\mathbf A_2 & \mathbf B_2 & \mathbf C_2 & \ddots & \vdots\\
\mathbf 0 & \mathbf A_3 & \mathbf B_3 & \ddots & \mathbf 0\\
\vdots & \ddots & \ddots & \ddots & \mathbf C_{2Q-1}\\
\mathbf 0 & \cdots & \mathbf 0 & \mathbf A_{2Q} & \mathbf B_{2Q}
\end{bmatrix},
\label{eq:S_block_tridiag}
\end{equation}
where $\mathbf B_\ell\in\mathbb C^{K\times K}$ are the diagonal blocks and
$\mathbf A_\ell,\mathbf C_\ell\in\mathbb C^{K\times K}$ are the lower and upper
off-diagonal blocks, respectively.
This block-tridiagonal structure is a direct consequence of the layered SIM
topology and of the dominant coupling pattern encoded in
$\bm S_{EE}^{(0)}$.

We seek $\mathbf T=\mathbf S^{-1}$, whose block entries
$\mathbf T_{\ell,r}\in\mathbb C^{K\times K}$ satisfy
\begin{equation}
\mathbf S\,\mathbf T = \mathbf I_{2QK}.
\label{eq:S_T_identity}
\end{equation}
The inverse $\mathbf T$ is computed by solving $2Q$ block linear systems,
each corresponding to one block-column of $\mathbf T$.
To this end, a block-Thomas factorization of $\mathbf S$ is first performed.

Define the forward Schur complements
\begin{equation}
\label{eq:block_thomas_forward_app}
\begin{aligned}
\mathbf F_1 &\triangleq \mathbf B_1,\\
\mathbf G_\ell &\triangleq \mathbf A_\ell\,\mathbf F_{\ell-1}^{-1},
\qquad \ell=2,\ldots,2Q,\\
\mathbf F_\ell &\triangleq \mathbf B_\ell-\mathbf G_\ell\,\mathbf C_{\ell-1},
\qquad \ell=2,\ldots,2Q.
\end{aligned}
\end{equation}
The recursion is well defined provided that all matrices
$\mathbf F_\ell$ are nonsingular.
The factorization in \eqref{eq:block_thomas_forward_app} is performed once
and reused for all block-columns of $\mathbf T$.

Let $\mathbf t^{(r)}$ denote the $r$-th block-column of $\mathbf T$,
\[
\mathbf t^{(r)}
=
\big[
\mathbf T_{1,r}^T,\,
\mathbf T_{2,r}^T,\,
\ldots,\,
\mathbf T_{2Q,r}^T
\big]^T,
\]
which solves
\begin{equation}
\mathbf S\,\mathbf t^{(r)} = \mathbf e^{(r)},
\label{eq:block_column_system}
\end{equation}
where $\mathbf e^{(r)}$ is the block vector whose $r$-th block equals
$\mathbf I_K$ and all other blocks are zero.

Define
\begin{equation}
\label{eq:block_forward_rhs_app}
\begin{aligned}
\mathbf d_1^{(r)} &\triangleq \mathbf e_1^{(r)},\\
\mathbf d_\ell^{(r)} &\triangleq
\mathbf e_\ell^{(r)}-\mathbf G_\ell\,\mathbf d_{\ell-1}^{(r)},
\qquad \ell=2,\ldots,2Q.
\end{aligned}
\end{equation}
The blocks of $\mathbf t^{(r)}$ are then obtained as
\begin{equation}
\label{eq:block_backward_app}
\begin{aligned}
\mathbf T_{2Q,r} &= \mathbf F_{2Q}^{-1}\,\mathbf d_{2Q}^{(r)},\\
\mathbf T_{\ell,r} &=
\mathbf F_\ell^{-1}
\Big(
\mathbf d_\ell^{(r)}-\mathbf C_\ell\,\mathbf T_{\ell+1,r}
\Big),
\qquad \ell=2Q-1,\ldots,1.
\end{aligned}
\end{equation}
Repeating the procedure
\eqref{eq:block_forward_rhs_app}–\eqref{eq:block_backward_app}
for $r=1,\ldots,2Q$ yields all block-columns of $\mathbf T$ and hence the
full inverse matrix.

\bibliographystyle{IEEEtran}
\bibliography{biblio_SIM}
\end{document}